\documentclass{aa}  

\usepackage{graphicx}
\usepackage{color}
\usepackage{enumerate}
\usepackage{subfig}
\usepackage{txfonts}
\usepackage{fontawesome}
\usepackage[]{hyperref}
\begin{document}

   \title{Discovering strongly lensed quasar candidates with catalogue-based methods from DESI Legacy Surveys}

   \author{Zizhao He
          \inst{1,2}
          \and
          Nan Li\inst{1}\fnmsep\thanks{nan.li@nao.cas.cn}
          \and
          Xiaoyue Cao\inst{2,3}
          \and
          Rui Li\inst{2,3}
          \and
          Hu Zou\inst{2,3}
          \and
          Simon Dye\inst{4}
          }
            \institute{Key lab of Space Astronomy and Technology, National Astronomical Observatories, 20A Datun Road, Chaoyang District, Beijing 100012, China; 
            %   \email{wuchterl@amok.ast.univie.ac.at}
            \and
		    School of Astronomy and Space Science, University of Chinese Academy of Sciences, Beijing 100049, China;
		    \and
            National Astronomical Observatories, Chinese Academy of Sciences, 20A Datun Road, Chaoyang District, Beijing 100012, China; 
            %  \email{c.ptolemy@hipparch.uheaven.space}
            %  \thanks{The university of heaven temporarily does not
                    %  accept e-mails}
            \and
            School of Physics and Astronomy, University of Nottingham, Nottingham, NG7 2RD, UK.
             }

   \date{Received XX, 2022; accepted XX, 2022}

% \abstract{}{}{}{}{} 
% 5 {} token are mandatory
 
  \abstract
  {The Hubble tension, revealed by a $\sim 5\sigma$ discrepancy between measurements of the Hubble-Lemaitre constant from early- and local-Universe observations, is one of the most significant problems in modern cosmology. In order to better understand the origin of this mismatch, independent techniques to measure $H_0$, such as strong lensing time delays, are required. Notably, the sample size of such systems is key to minimising statistical uncertainties and cosmic variance, which can be improved by exploring the datasets of large-scale sky surveys like DESI (Dark Energy Spectroscopic Instrument).}
  {We identify possible strong lensing time-delay systems within DESI by selecting candidate multiply imaged lensed quasars from a catalogue of 24,440,816 candidate QSOs contained in the 9th data release of the DESI Legacy Imaging Surveys (DESI-LS).}
  {Using a friend-of-friends-like algorithm on spatial co-ordinates, our method generates an initial list of compact quasar groups. This list is subsequently filtered using a measure of the similarity of colours of a group's members and the likelihood that they are quasars. A visual inspection finally selects candidate strong lensing systems based on the spatial configuration of the group members.}
  {We identify 620 new candidate multiply imaged lensed quasars  (101 Grade-A, 214 Grade-B, 305 Grade-C). This number excludes 53 known spectroscopically confirmed systems and existing candidate systems identified in other similar catalogues. When available, these new candidates will be further checked by combining the spectroscopic and photometric data from DESI. The catalogues and images of the candidates in this work are available online\thanks{\url{https://github.com/EigenHermit/lensed_qso_cand_catalogue_He-22/}}.} 
  {}
% \footnote{\url{https://github.com/EigenHermit/lensed_qso_cand_catalogue_He-22/}}

   \maketitle

%-------------------------------------------------------------------
% \faGithub\href{https://github.com/EigenHermit/lensed_qso_cand_catalogue_He-22/}{~here.} 

\section{Introduction}
Measurements of the Hubble-Lemaitre constant ($H_0$) from the local and the early Universe have shown substantial differences \citep{Riess2019,Verde2019}, for instance, measurements of the cosmic microwave background \citep[CMB; see][]{Bennett2013,Planck_Collaboration2020} and baryon acoustic oscillations \citep[BAO; see][]{Addison2018,Abbott2019} and those made in the more local Universe using supernovae \citep[SNe; see][]{Dhawan2018,Macaulay2019}, the tip of the red giant branch \citep[TRGB; see][]{Freedman2019,Yuan2019}, and Cepheid variables \citep[][]{Pietrzynski2019,Riess2019}. This so-called `Hubble tension' is considered one of the most significant crises of modern cosmology. Many efforts have been devoted to solving the problem \citep{Vagnozzi2020, Adhikari2022, Goicoechea2022, Niedermann2022}, but the answer remains inconclusive. 

Independent of all of the aforementioned methods, strong lensing time delays provide valuable measurements of $H_0$ \citep{Birrer2019, Liao2019, Wong2020, Shajib2020} that may assist in understanding these discrepancies once potential selection bias and unignorable statistical uncertainties in the technique are fully calibrated. Achieving this will require a much larger sample of time delay systems than what is currently available \citep{Shajib2018} via the gravitationally lensed QSO database\footnote{\url{https://research.ast.cam.ac.uk/lensedquasars/index.html}} (GLQ).

In addition to constraining the cosmological model, strong lensing time delay systems, typically multiply imaged lensed QSOs, provide valuable insight to astrophysical problems such as constraining distributions of dark and luminous matter of the lenses \citep{Oguri2014, Suyu2014, Sonnenfeld2021, Vyvere2021}, and uncovering the properties of distant active galactic nuclei (AGN) and their host galaxies to a level of detail not possible without lens magnification \citep[e.g.,][]{McGreer2010, More2015, Fan2019, Yue2021}. In the case of the latter, microlensing caused by small structures within the lens have revealed fine-level details of AGN morphology such as accretion disk characteristics \citep{Anguita2008, Motta2012, Braibant2014, Fian2021}. 

With increasing depth and sky coverage of large-scale surveys, the sample size of multiply imaged QSOs is predicted to grow remarkably. For instance, within third generation surveys like the Dark Energy Survey \citep[DES,][]{Akhazhanov2022}, KiDS \citep[Kilo-Degree Survey,][]{Kuijken2019}, and Gaia \citep{lemon2022}, there are an expected $\sim 2000$ multiply imaged QSOs systems \citep{Oguri2010}. The current mainstream proven strategy for seeking these lensed QSOs comprises two steps: 1) finding candidates with high completeness and 2) confirming the candidates with spectra to improve the purity. Previous studies have seen construction of several candidate catalogues, such as those by \citep{Agnello2015, Martins2018, Spiniello2018, Spiniello2019, Wu2022, Akhazhanov2022} and subsequent spectroscopic follow-up has provided a number of confirmed lensed QSO samples \citep[see, e.g. ][]{Lemon2018, Lemon2019}. Sample sizes are set to increase even further with upcoming fourth-generation surveys like Euclid \citep{Euclid-intro}, the Large Survey of Space and Time \citep[LSST][]{LSST-intro}, Roman \citep{Roman-intro}, and the Chinese Space Station Telescope \citep[CSST;][]{CSST-intro}.

In this paper, we build a catalogue-based algorithm for finding the candidates of multiply imaged lensed QSOs from the QSO candidate catalogue of \citet{He2022} extracted from the 9th data release of the Dark Energy Spectroscopic Instrument Legacy Imaging Surveys \citep[DESI-LS;][]{Dey2019} which covers $\sim 20,000$\,deg$^2$ of the extragalactic sky in $g,r,z$-bands. Specifically, we compile a catalogue containing 971 multiply imaged lensed systems, 620 of which are new candidates not contained in the known lensed QSO databases of \citet[][D22 hereafter]{Dawes2022} or \citet[][L22 hereafter]{lemon2022}. Catalogues and images are made publicly available online \footnote{\url{https://github.com/EigenHermit/lensed\_qso\_cand\_catalogue\_He-22}}.
   
This paper is organised as follows. Sec.\,\ref{sec:data} introduces the datasets used in this project. The methodology for identifying strongly lensed QSO candidates is detailed in Sec.\,\ref{sec:method}. The results, including candidate catalogues, the images and lens models of selected candidates, and the comparison with known lensed QSOs and other works are presented in Sec.\ref{sec:res}. Finally, Sec.\,\ref{sec:conclusion} delivers the discussion and conclusions. In this paper, a fiducial cosmological model $\Omega_m=0.26$, $\Omega_{DE}=0.74$, $h=0.72$, $w_0=-1$ and $w_a=0$ is assumed, matching that adopted by \citet[][OM10 hereafter]{Oguri2010}. Unless otherwise stated, all magnitudes quoted in this paper are in the AB system.
%--------------------------------------------------------------------
\section{Datasets}
\label{sec:data}

In this section we describe the QSO candidate catalogue, which we mine for multiply imaged QSO candidates, and a reference sample of known lensed QSOs, which we use for optimising the mining process. The former is introduced in Sec.\,\ref{sec:qcc}, and the latter in Sec.\,\ref{sec:tst_sample}.

\subsection{The QSO candidate catalogue}
\label{sec:qcc}

\begin{figure*}
         \centering
         \includegraphics[scale=0.35]{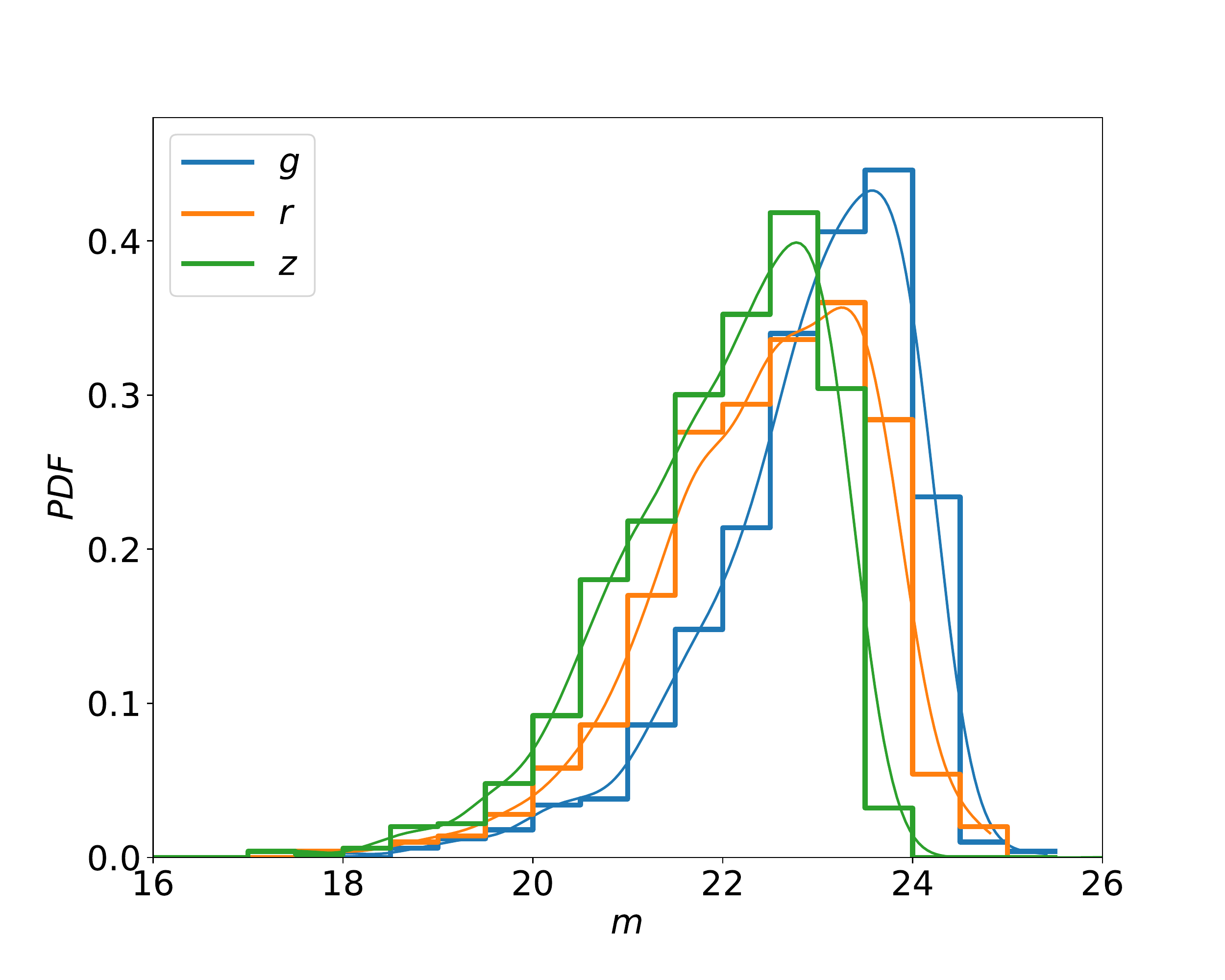}
         \includegraphics[scale=0.35]{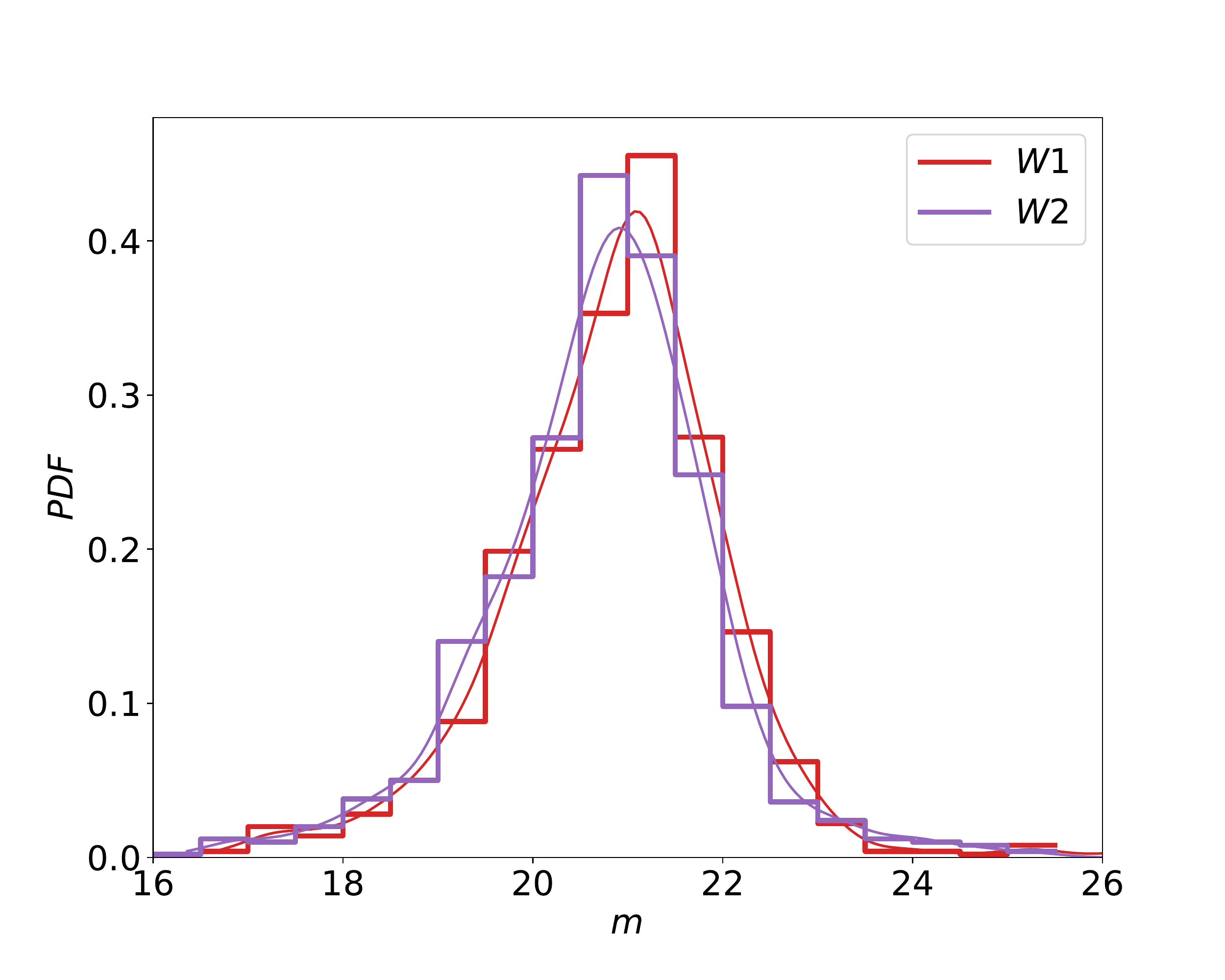}
         \caption{Magnitude distributions of the QCC in $g,r,z$ (left panel) and $W1,W2$ (right panel). The kernel density estimate curves (solid lines) are generated by {\tt kdeplot} in {\tt seaborn} package with $binsize=0.5$ and the default Gaussian kernel.}
         \label{fig:dist_qcc}
    \end{figure*}

The QSO candidate catalogue (QCC) of \cite{He2022} contains $24,440,816$ objects in total. It was created from the point-like sources identified in DESI-LS DR9 using a Random Forest \citep[RF;][]{Breiman2001} classification model. The catalogue includes the RA, Dec, the five band magnitudes ($g$, $r$, $z$, $W1$, $W2$), and the probability of being a QSO given by the RF model. The magnitude distributions are shown in Fig.\,\ref{fig:dist_qcc}. The $r$-band magnitude of the QCC ranges from 18 to 26, with a mean of 22.44. Evaluated with the testing set that mimics the magnitude and colour distributions of the point-like sources of DESI-LS, the candidate catalogue's completeness and purity are $\sim 99\%$ and $\sim 25\%$, respectively. 

DESI-LS covers an area of $\sim 20,000$ deg$^2$ in $g,r,z$ and comprises three different sub-projects: DECaLS\footnote{\url{https://www.legacysurvey.org/decamls/}} \citep[The Dark Energy Camera Legacy Survey,][]{Dey2019}, BASS\footnote{\url{ https://www.legacysurvey.org/bass/}} \citep[Beijing-Arizona Sky Survey,][]{Zou2017,Zou2019} and MzLS\footnote{\url{https://www.legacysurvey.org/mzls/}} \citep[The Mayall $z$-band Legacy Survey,][]{Dey2019}. The point source sensitivities of these surveys are as follows: for DECaLS, the 5$\sigma$ detection limits in AB mag for a point source in individual images are 23.95, 23.54, and 22.50 of $g,r,z$-bands; for BASS, they are 23.65 ($g$-band) and 23.08 ($r$-band); for MzLS, it is 22.60 ($z$-band). The DESI-LS DR9 catalogue also includes four mid-infrared bands at 3.4, 4.6, 12, and 22 $\mu m$ (corresponding to $W1$, $W2$, $W3$ and $W4$ respectively) observed by the Wide-field Infrared Survey Explorer\footnote{\url{http://wise.ssl.berkeley.edu/index.html}} \citep[WISE, ][]{Wright2010}.

\subsection{Reference sample}
\label{sec:tst_sample}

    We adopt a reference sample containing 57 spectroscopically confirmed multiply imaged QSO systems to optimise the candidate selection process, such as fine-tuning selection thresholds and training the inspectors. The reference sample (referred to as RLQ hereafter) is acquired by cross-matching the GLQ and QCC. There are 111 objects ($\sim 50.5\%$ of all objects in GLQ, composed of 93 pairs + 18 quads) having at least one matched member and 57 of these 111 objects have at least two matched members in the QCC. The presence of multiple matched members is an indication that our group finder can successfully detect all of these multiply imaged QSO systems (see Sec.\,\ref{sec:group_finder} for more details).
    
    % "Multiple matched members" means that our group finder can successfully detect all these multiply-imaged QSO systems theoretically (see Sec.\,\ref{sec:group_finder} for more details).

    % Multiple matched members means that our group finder successfully detects multiply imaged QSO systems
    
\begin{figure*}
        \centering
        \includegraphics[scale=0.355]{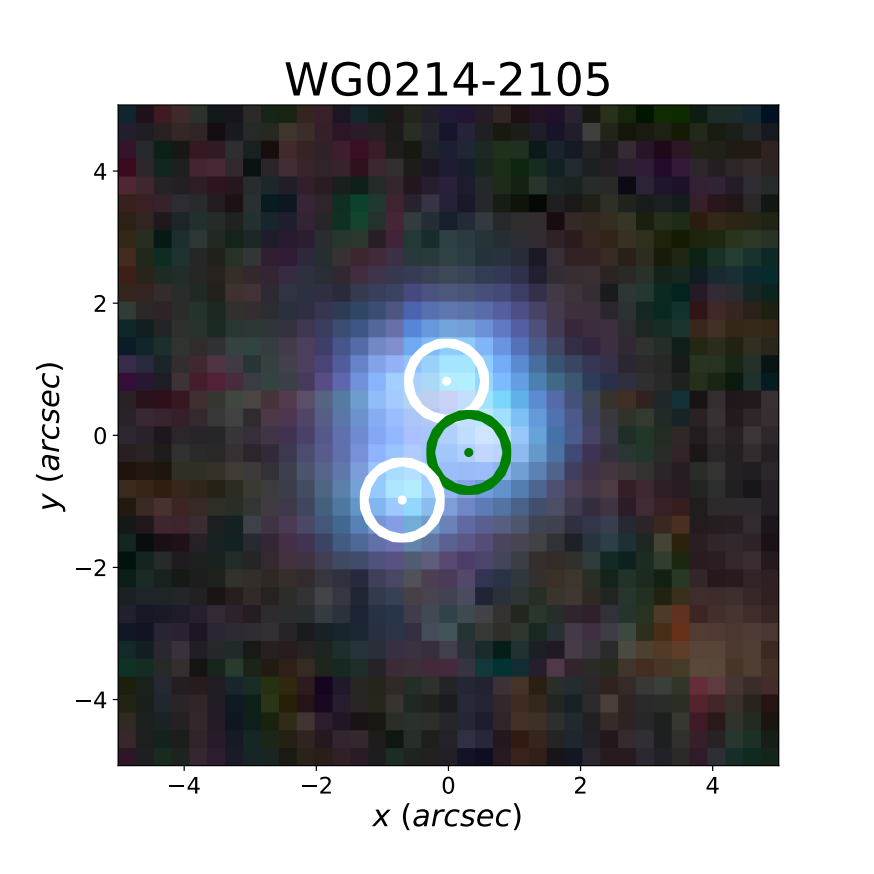}
        \includegraphics[scale=0.29]{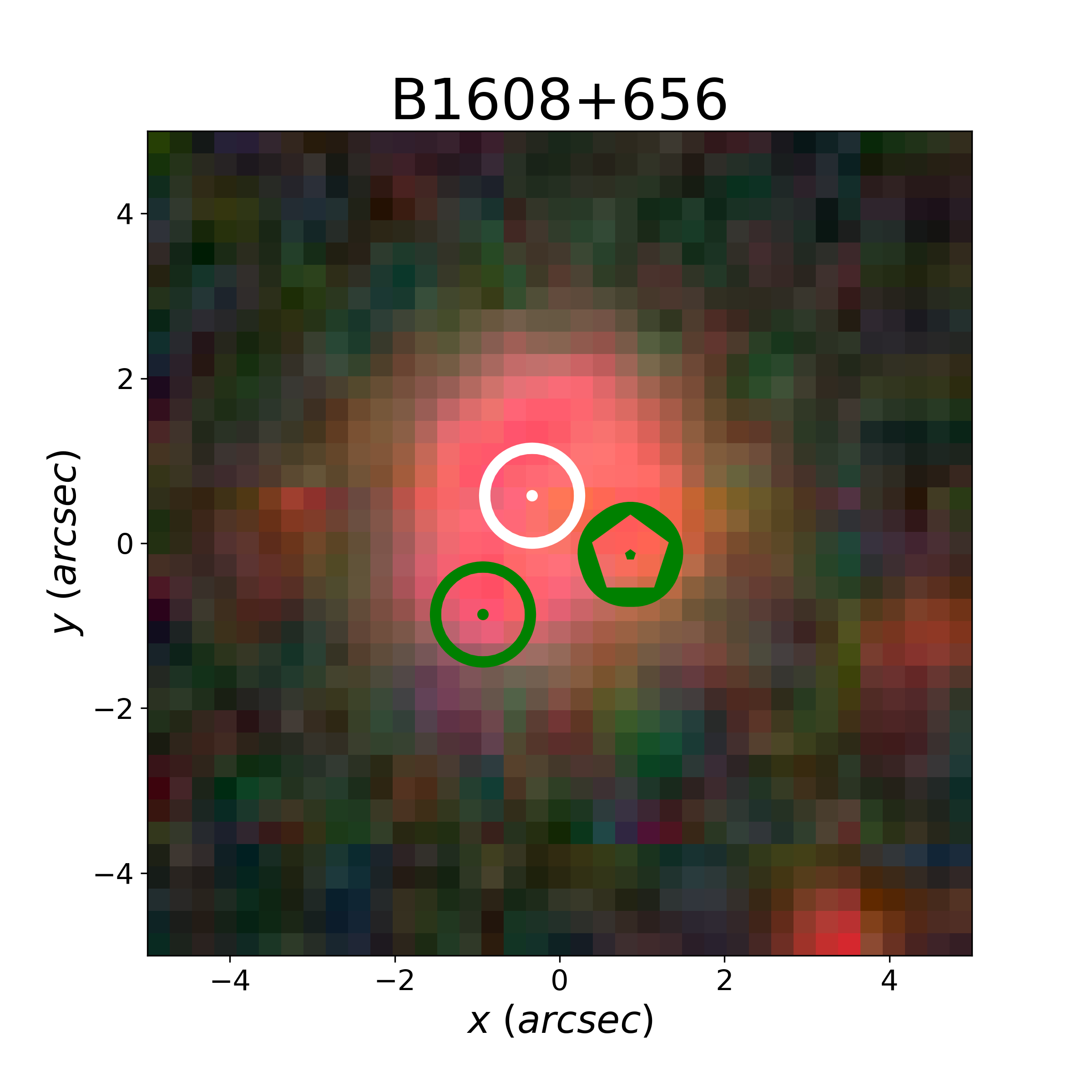}
        \includegraphics[scale=0.355]{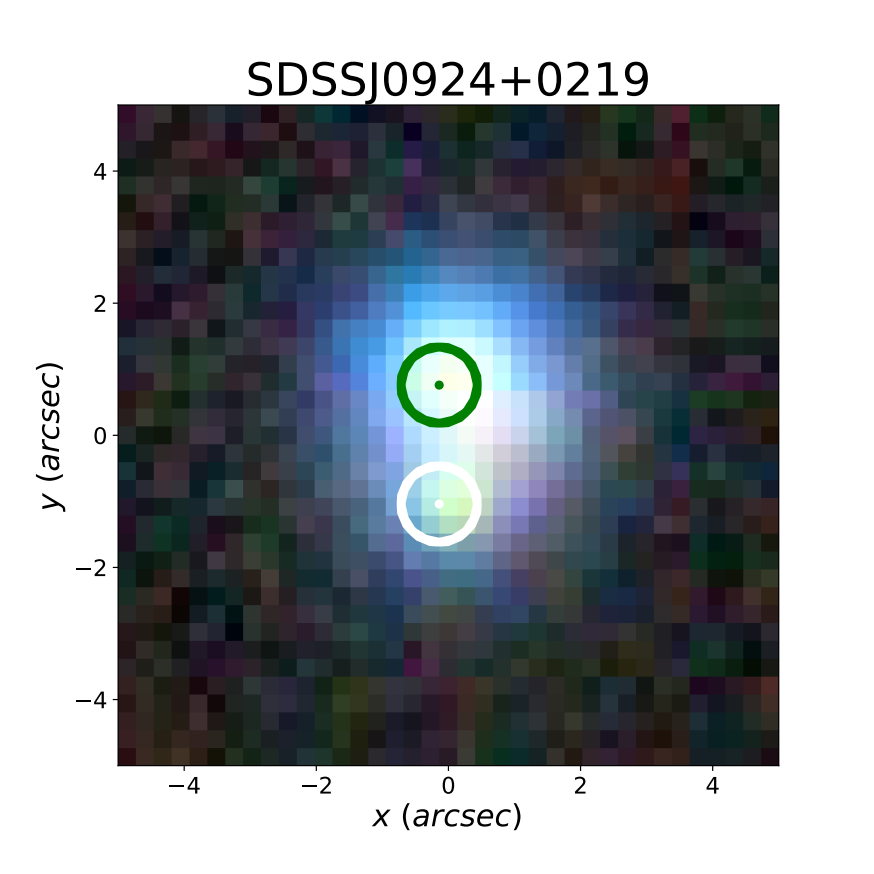}
        \caption{Demonstrating the three possible situations that cause a known multiply imaged QSO to have only one match in the QCC. The colours of the cycles represent the morphology types in DESI-LS catalogues. Green circles represent point-like sources whereas white circles represent extended sources. For WG0214-2105, three out of its four images are recorded in the catalogue, but only one is labelled as being point-like. For B1608+656, the image with the green pentagon marker lacks a $g$-band detection. In the case of SDSSJ0924+0219, three of the lensed images have been identified as a single point-like source, and the fourth image has been identified as an extended object.}
        \label{fig:demo_of_miss}
    \end{figure*}

    Our reference sample comprises only those objects with two or more matches. The remaining excluded 54 systems with only one match arise due to three possible situations:
    \begin{enumerate}
        \item In a given multiply imaged QSO system, only one of the lensed images is labelled as the point-like source in the DESI-LS DR9 catalogue, and the other lensed images are labelled as extended sources. Hence the QCC did not include these multiple images in the first place (see the left panel of Fig.\,\ref{fig:demo_of_miss}).
        \item As shown in the middle panel of Fig.\,\ref{fig:demo_of_miss}, not all of the multiple images of lensed QSOs are detected in all five-bands  and thus are not included in the QCC.
        \item The multiple images are smeared by the imaging point spread function (PSF) and therefore have been labelled as a single point-like source in the DESI-LS DR9 catalogue and so too in the QCC (see the right panel of Fig.\,\ref{fig:demo_of_miss}). 
    \end{enumerate}

    The catalogue of RLQ holds both the information provided in GLQ (including RA, Dec, Name, lens and source redshifts, and the number of images) and the labels and scores attributed in this work, such as the internal IDs given by the group finder, $S_{colour}$ and $S_{RF}$ (defined in Sec.\,\ref{sec:select_groups_bycolor} and \ref{sec:select_groups_byRF}), and labels indicating whether the system exists in the sample for visual inspection (Sec.\,\ref{sec:vs}). The detailed RLQ catalogue and the corresponding description are published online \footnote{\url{https://github.com/EigenHermit/lensed\_qso\_cand\_catalogue\_He-22/blob/main/RLQ/RLQ.csv}}  \footnote{\url{https://github.com/EigenHermit/lensed\_qso\_cand\_catalogue\_He-22/blob/main/RLQ/description}}. 

\section{Methodology}
\label{sec:method}
    
    Our procedure for mining multiply imaged lensed QSOs in the QCC is broadly divided into three phases. Sec.\,\ref{sec:group_finder} describes the first phase, where we use a grid-search algorithm on the angular positions of the members in the QCC to identify goups of candidate QSOs. The second phase cleans these groups depending on whether they contain two (Sec.\,\ref{sec:select_pair}) or more (Sec.\,\ref{sec:select_mul}) members. The third phase applies visual inspection to finalise the list of multiply imaged candidates (Sec.\,\ref{sec:vs}). 
    
\subsection{QSO group finder}
\label{sec:group_finder} 
Our QSO group finder selects QSO candidate groups within the QCC based on their angular positions. This has four steps: 
    \begin{enumerate}
        \item All sources in the QCC are put into grids generated by {\tt HEALPix} \citep{Gorski1999} with nside=$2^{15}$, giving rise to a grid bin size of $6.4 \times 6.4$ arcsec$^2$. The $24,440,816$ sources in the QCC occupy $23,394,395$ HEALPix grids.
        \item The program begins with grid bins that contain at least one QSO candidate. For each of these bins, the surrounding eight grid bins are checked for QSO candidates. Surrounding bins that contain QSOs are then connected to the central bin. After that, the surrounding bins of new groups are checked, and if there are QSO candidates in the neighbouring pixels of all the members of the group, the new candidates are further added to the group. This process is then repeated until there are no more candidates found in the surrounding bins of the final group.
        %and it cycles to check all neighbour grids until there are no more candidates in all neighbour grids of the group;
        \item Once a group of bins is obtained in this way, the QSO candidates in the group are labelled as belonging to that QSO candidate group.
        \item For a QSO candidate group holding two members only, their angular distance must be greater than $0.5$ but less than $10$ arcsec, otherwise, the group is removed.
    \end{enumerate}

    As a result, 562,206 QSO candidate groups are identified and placed into a candidate QSO group catalogue (QGC). As expected, the QGC includes all groups contained in the RLQ. The QGC provides the RA, Dec, the probability of being a QSO given by \cite{He2022}, and the five-band magnitudes $g, r, z, W1, W2$ for each candidate of every group. (The QGC also gives the quantities $S_{colour}$ and $S_{RF}$ which we define below.) 
    
    We divide the QGC into two parts, one part containing groups with two members and the other containing groups with more than two members. We label these PAIR and MUL respectively. PAIR includes 462,608 systems, while MUL includes 99,538 systems. We also label those groups found in the RLQ as either RLQ-PAIR (47 systems) or RLQ-MUL (10 systems) depending on whether they belong to PAIR or MUL. The QGC and its corresponding description is published online\footnote{\url{https://github.com/EigenHermit/lensed\_qso\_cand\_catalogue\_He-22/tree/main/QGC}}.

\subsection{QSO candidate groups in PAIR}
\label{sec:select_pair}

    PAIRs are approximately five times more prevalent in the  QGC than MULs. We discard all PAIRs that have large colour differences (see below) and/or low probabilities of their members being QSO candidates. As we discuss in Sec.\,\ref{sec:select_mul}, we implement a different cleaning strategy for the MULs.

\subsubsection{Selection with colour similarity}
\label{sec:select_groups_bycolor}
    Since gravitational lensing is an achromatic process, differences in colour between multiple images can only arise through differing scattering processes along the different paths between the observer and source. Unless a lens has strong dust gradients, the colours of lensed images are therefore generally very similar within a given multiply imaged system. As such, we can use colour similarity to eliminate PAIRs that are not likely double-image systems. Hence, we define the colour similarity of a group in PAIR, $S_{colour}$, as:
    \begin{equation}
        S_{colour}=
        \begin{cases}
        1 - \frac{1}{10}\sum_{i=1}^{10}\sigma_{i}& {\rm if} \,\,\,\, {\frac{1}{10}\sum_{i=1}^{10}\sigma_{i}<1}\\
        0& {\rm if} \,\,\,\, {\frac{1}{10}\sum_{i=1}^{10}\sigma_{i}\geq1}
        \end{cases},
        \label{eq:s_colour}
    \end{equation}
    where $\sigma_{i}$ is the standard deviation of the $i$th colour (computed over both members of the PAIR) out of the 10 unique colours provided by the $g, r, z, W1$ and $W2$ magnitudes in the QCC.

    All PAIRs with a value of $S_{colour}$ less a given threshold are rejected. We choose $0.5$ as the threshold in this work as a compromise between rejecting as many systems in the QGC as possible whilst minimising the rejection of confirmed RLQ-PAIR systems. This choice of threshold rejects $\sim 36\%$ of systems in the QGC while only losing $\sim 4\%$ of confirmed systems classified as RLQ-PAIR. We label the remaining 297,502 groups of QSO PAIR candidates as PAIR-CS. Two known lensed QSOs are missing: RXJ0911+0551 and SBS1520+530. These systems exhibit significantly different colours between their images, presumably due to very different levels of dust encountered along the path to each image. In particular, these systems have significant differences between images in the colours that include the $W1$ and $W2$ magnitudes. However, if we do not include these magnitudes, the rejection rate drops dramatically which leads to a much higher workload for human inspection. We discuss this further in Sec.\,\ref{sec:conclusion}.

\subsubsection{Selection with the probabilities of being QSOs}
\label{sec:select_groups_byRF}
    
        We use an additional metric, $S_{RF}$, that measures the overall probability of the candidates truly being QSOs, to remove the groups with possible fake QSOs. For a given QSO candidate group, this score is defined as:
    \begin{equation}
        S_{RF} = \frac{1}{n}\sum_{i=1}^{n}prob_i
        \label{eq:s_rf}
    \end{equation}
    where $n$ is the number of QSO candidates in the group. The quantity $prob\_i$ is provided in the QCC and gives the probability of a candidate being a true QSO. The score $S_{RF}$ therefore represents the likelihood that a candidate group is a QSO group.

    In this work, we choose a threshold of $S_{RF}=0.85$ and reject any groups in PAIR-CS with a value of $S_{RF}$ less than this. This removes two more confirmed multiply imaged QSOs (SBS0909+532 and DESJ0405-3308) leaving 43 out of the initial 47 confirmed systems in RLQ-PAIR. Overall, 102,468 ($\sim $ 22.2\% PAIR, $\sim $ 34.4\% PAIR-CS) groups in PAIR are selected and labelled as PAIR-CS-RF. 

    Since we provide both $S_{RF}$ and $S_{colour}$ in our online catalogue, users of the data can apply different thresholds according to whether a higher recall rate or a better precision is required.
    
    \subsection{QSO candidate groups in MUL}
    \label{sec:select_mul}

    Our procedure for cleaning PAIRs is not appropriate for candidate groups classified as MUL since it is overly sensitive to groups that contain additional contaminating members. As such, we define a different strategy for the MUL groups in the QGC.
    
    First, for each group, members with $S_{RF}<0.85$ are removed. Then, groups containing only one member after the above process are also discarded. The remaining 52,582 groups ($\sim 52.8\%$ MUL) are labelled MUL-RF. Collectively, groups within MUL-RF contain a total of 134,976 QSO candidate members. Secondly, we explore the colour similarity of the members in each group to quantify the possibility of each group being a multiply imaged QSO system. The details of this step are as follows: 

    \begin{enumerate}
        \item For a given group, $S_{colour}$ is calculated for all possible combinations of group members, from those containing two members to the largest combination that contains all members.
        \item All combinations with $S_{colour} < 0.5$ are rejected. 
        \item Finally, the combination containing the most members (and that with the highest $S_{colour}$ if there is more than one of these) is retained, and the others are rejected. 
    \end{enumerate}

As a result, 45,905 groups (containing a total of 111,761 members) remain ($\sim 46.1\%$ MUL, $\sim 87.3\%$ MUL-RF) and are labelled MUL-RF-CS. In MUL-RF-CS, all ten of the confirmed systems in RLQ-MUL are retained. Together, the MUL-RF-CS and PAIR-CS-RF contain 53 out of 57 RLQs; i.e., the recovery rate is $\sim 93\%$.
    
    \subsection{Visual inspection}
    \label{sec:vs}
    
    The final phase of our mining procedure implements human visual inspection (VI) to improve the true positive rate and to grade the multiply imaged QSOs candidates. All inspectors are initially trained by visually analysing the images of the 53 confirmed lensed QSOs in PAIR-CS-RF and MUL-RF-CS. Then, all candidates in PAIR-CS-RF and MUL-RF-CS are inspected, yielding a total of 971 candidates. These 971 candidates are then graded into one of three grades, A, B, or C by two inspectors independently (the first and fourth author). An A grade is awarded to the most reliable systems with strong lensing features similar to the confirmed systems. B grades are awarded to systems with features that are less similar to the confirmed systems, such as larger image separation. To help the inspectors improve their objectivity, the statistics $S_{RF}$ and $S_{colour}$ are referred to. Grade C systems typically have a few lensing features; if they are lensed QSOs, their configurations are atypical.
    
    The grading process is naturally subjective, so the grades of the two inspectors are listed simultaneously in our online catalogue to give an indication of human bias. For simplicity, the grading results presented in this paper are solely those given by the first author unless otherwise stated. The following features are considered during the grading: 
    \begin{itemize}
        \item the higher $S_{colour}$ and $S_{RF}$, the better;
        \item the existence of an apparent lens galaxy makes the system more plausible, especially if it has a red colour;
        \item separations between images on opposite sides of the lens should usually be less than 3 arcsec, and a $[0.5,1.5]$ arcsec separation makes a candidate more convincing;
        \item for the pairs with apparent lens light, the angle between the two position vectors (measured from the lens centre to the image) is greater than 120 degrees \citep{Chan2015}. The candidates that do not satisfy this criterion are rejected or graded as C;
        \item for the pairs without apparent lens light, the $S_{colour}$ and $S_{RF}$ influence the grading more strongly;
        \item for quads, the configuration of the multiple images should be similar to those of the systems in the RLQ.
    \end{itemize}

    The resulting visually inspected catalogues are labelled PAIR-CS-RF-VI and MUL-RF-CS-VI. We combine the 971 candidates contained in total by these two catalogues into a single catalogue that we refer to as H22 hereafter. This catalogue is available online\footnote{\url{https://github.com/EigenHermit/lensed\_qso\_cand\_catalogue\_He-22/blob/main/Candidates/Catalogues/H22.csv}}.

\section{Results}
\label{sec:res}
    
    In this Section, we present the statistical properties of the candidates, comparing our final H22 candidate catalogue with existing known multiply imaged QSOs and two candidate catalogues determined by other studies.

    \subsection{Lensed QSOs candidates in this work}
    \label{sec:all_cand}

    Table \ref{tab:all_cand} lists the properties given for every candidate lensed QSO system in our H22 catalogue. Included in this list of properties is an internal ID linking to the QGC, the RA and Dec of the candidate,  the awarded grade, the evaluation metrics ($S_{RF}$, $S_{colour}$) and physical parameters such as image separation and number of images. Systems that have been previously modelled also contain the axis ratio, position angle and Einstein radius lens model parameters. We also include the labels `in\_L22' and `in\_D22' that indicate overlap with the candidate catalogues of \citet{lemon2022} and \citet{Dawes2022} respectively. Where an overlap occurs, we give the grade/classification awarded by that catalogue in the properties `Classification\_by\_L22' and `Grade\_by\_D22'.

    We also provide an additional catalogue which we refer to as H22-details, that includes all images belonging to each candidate lensed QSO system in H22\footnote{\url{https://github.com/EigenHermit/lensed_qso_cand_catalogue_He-22/blob/main/Candidates/Catalogues/H22-details.csv}}). Table \ref{tab:cand_details} lists the properties given for each image of each candidate lensed QSO system. These include the internal ID that links to the QGC (different images of one candidate share the same ID), RA, Dec, $g$, $r$, $z$, $W1$, $W2$ and the source redshift (if applicable) from the Sloan Digital Sky Survey \citep[SDSS,][]{Blanton2017} extended Baryon Oscillation Spectroscopic Survey \citep[eBOSS, ][]{Dawson2016} DR16. The RA and Dec here are the coordinates of the images in candidate systems.

\begin{figure}
        \centering
        \includegraphics[scale=0.45]{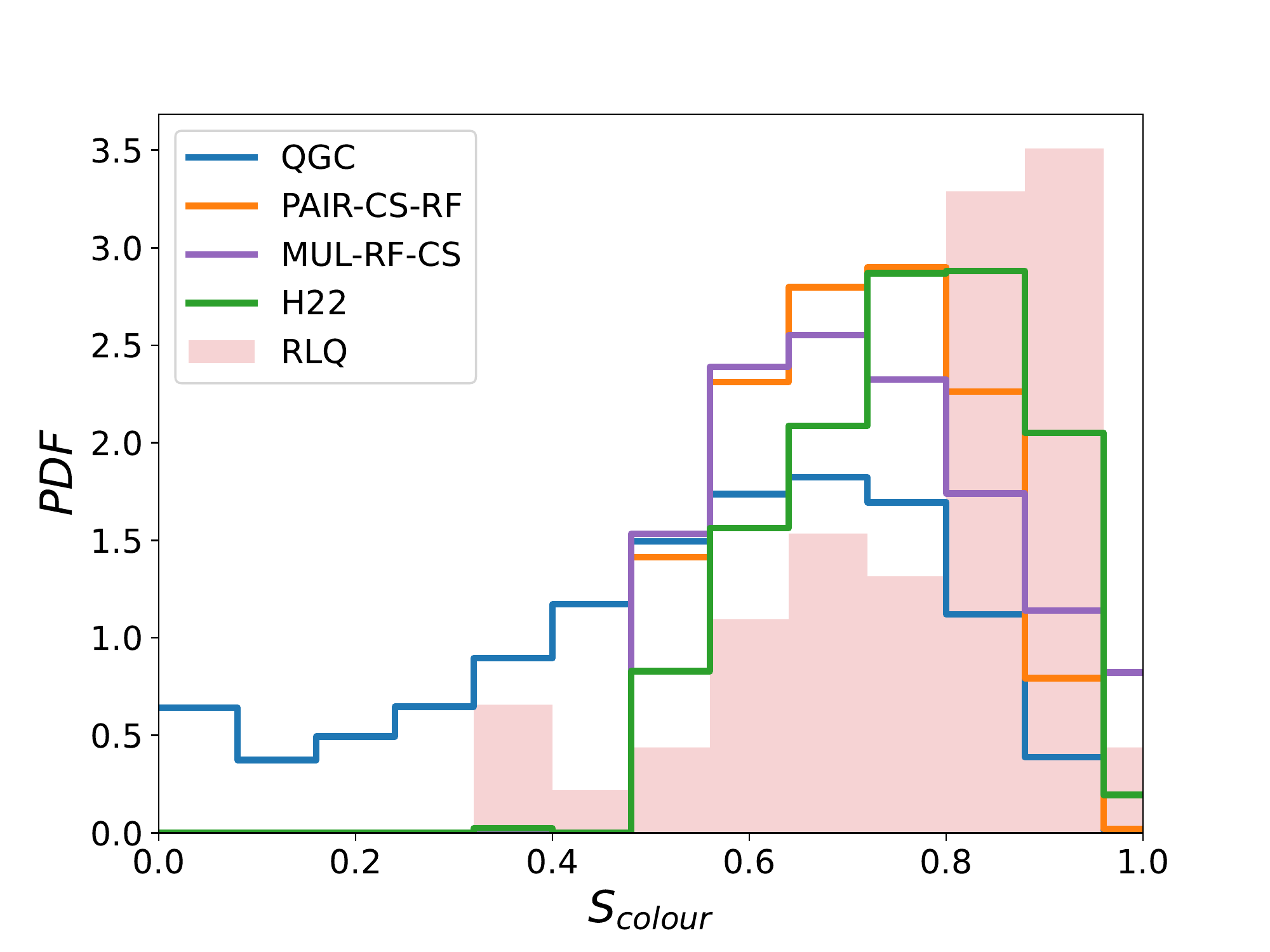}
        \includegraphics[scale=0.45]{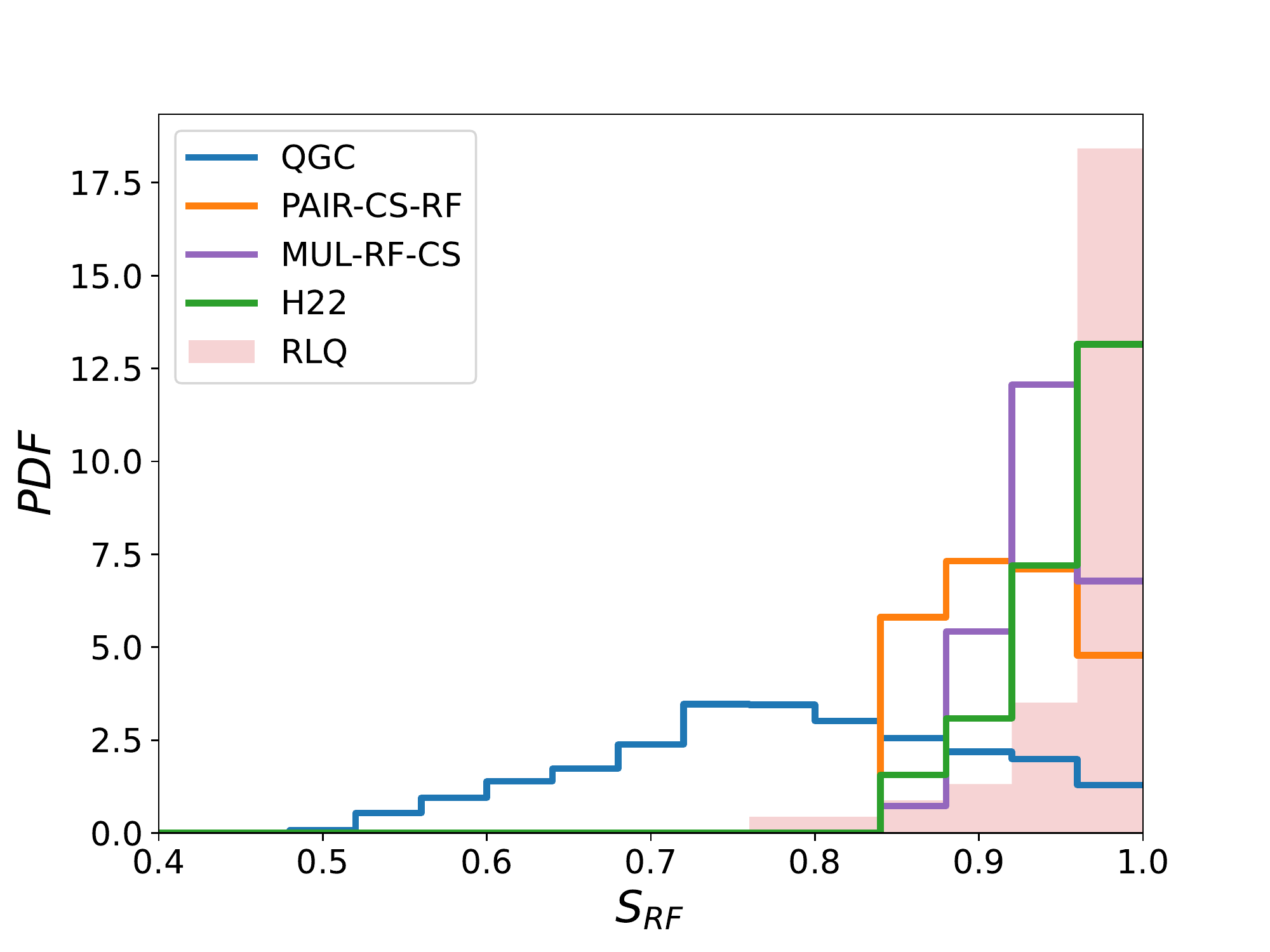}
        \caption{The distributions of $S_{colour}$ (upper panel) and $S_{RF}$ (lower panel) for systems in QGC, PAIR-CS-RF, MUL-RF-CS, H22 (all candidates), and RLQ.}
        \label{fig:s_dist}
    \end{figure}

    The distributions of $S_{colour}$ and $S_{RF}$ are plotted in Fig.\,\ref{fig:s_dist} for H22, the QGC, PAIR-CS-RF, MUL-RF-CS, and the catalogue of confirmed systems, RLQ. A general observation is that the distributions of PAIR-CS-RF, MUL-RF-CS and H22 match those of the RLQ catalogue significantly better than the QGC distributions, although this is to be expected given the selection criteria we have applied based on $S_{colour}$ and $S_{RF}$. In addition, the difference between the RLQ and QGC catalogues is much more pronounced for $S_{RF}$, indicating that this is a more efficient statistic on which to select lensed QSOs. It is also apparent from Fig.\,\ref{fig:s_dist} that H22 more closely matches the distributions of RLQ than PAIR-CS-RF or MUL-RF-CS do, demonstrating that the final step of human visual inspection has made a significant improvement. However, as is mentioned in Sec.\,\ref{sec:method}, our catalogue-based approach may miss the candidates that do not exist in the parent samples (the QCC), if also considering the various selection criteria for compiling RLQ, the difference between H22 and RLQ is inevitable. This difference can be reduced when spectroscopic follow-ups remove the false positives in H22.

\begin{figure*}
        \centering
        \centerline{\textbf{Example systems in both H22 and RLQ}}
        \includegraphics[scale=0.29]{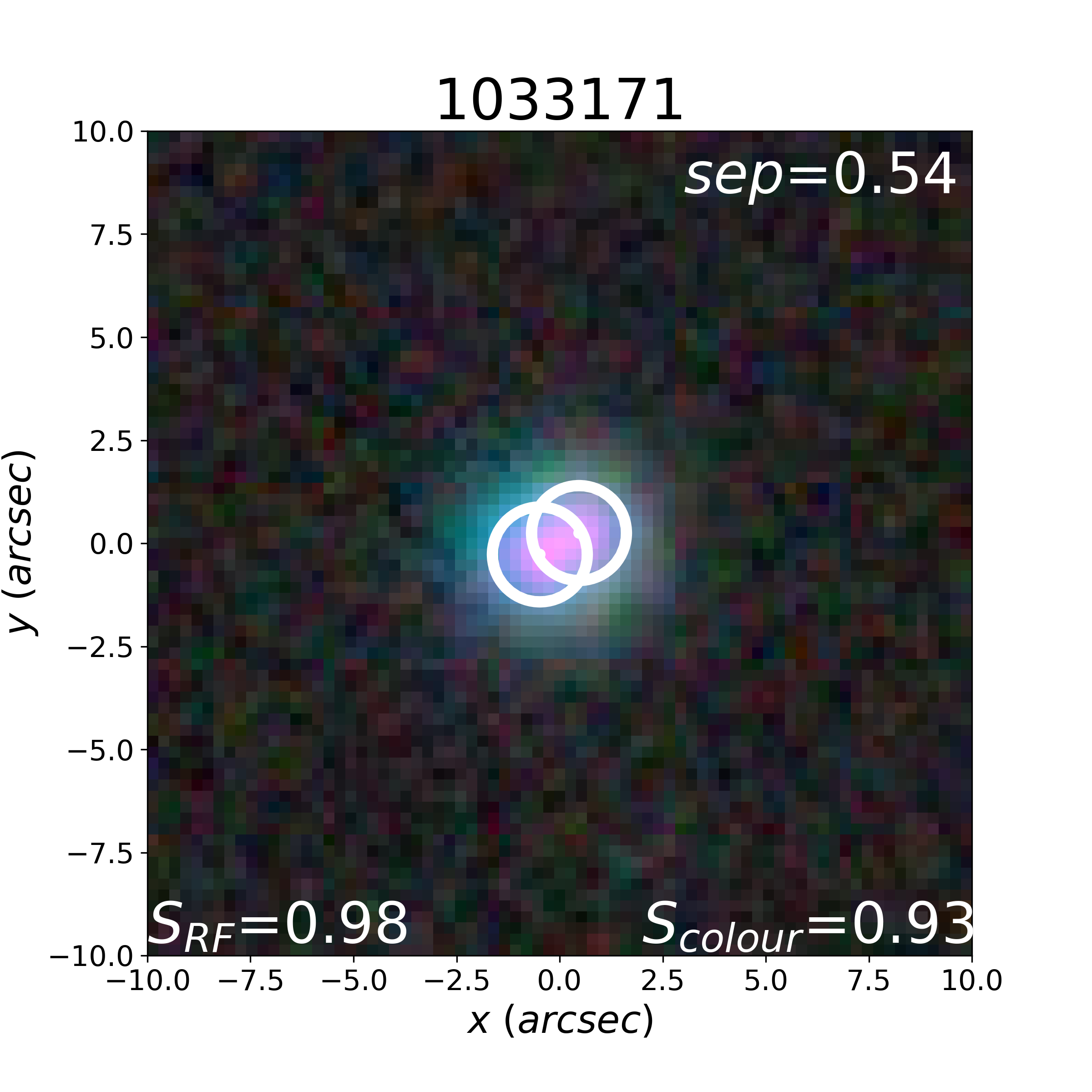}
        \includegraphics[scale=0.29]{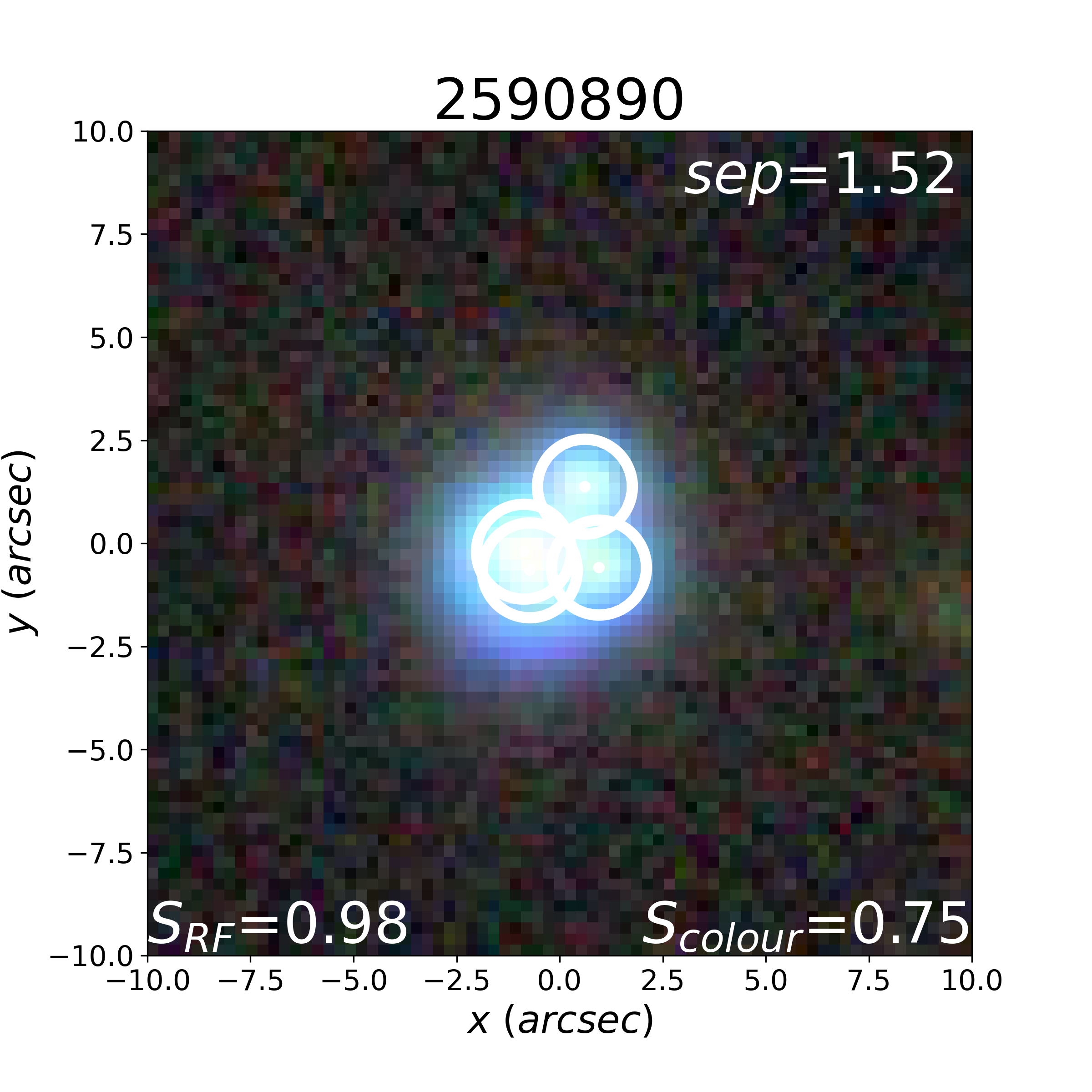}
        \includegraphics[scale=0.29]{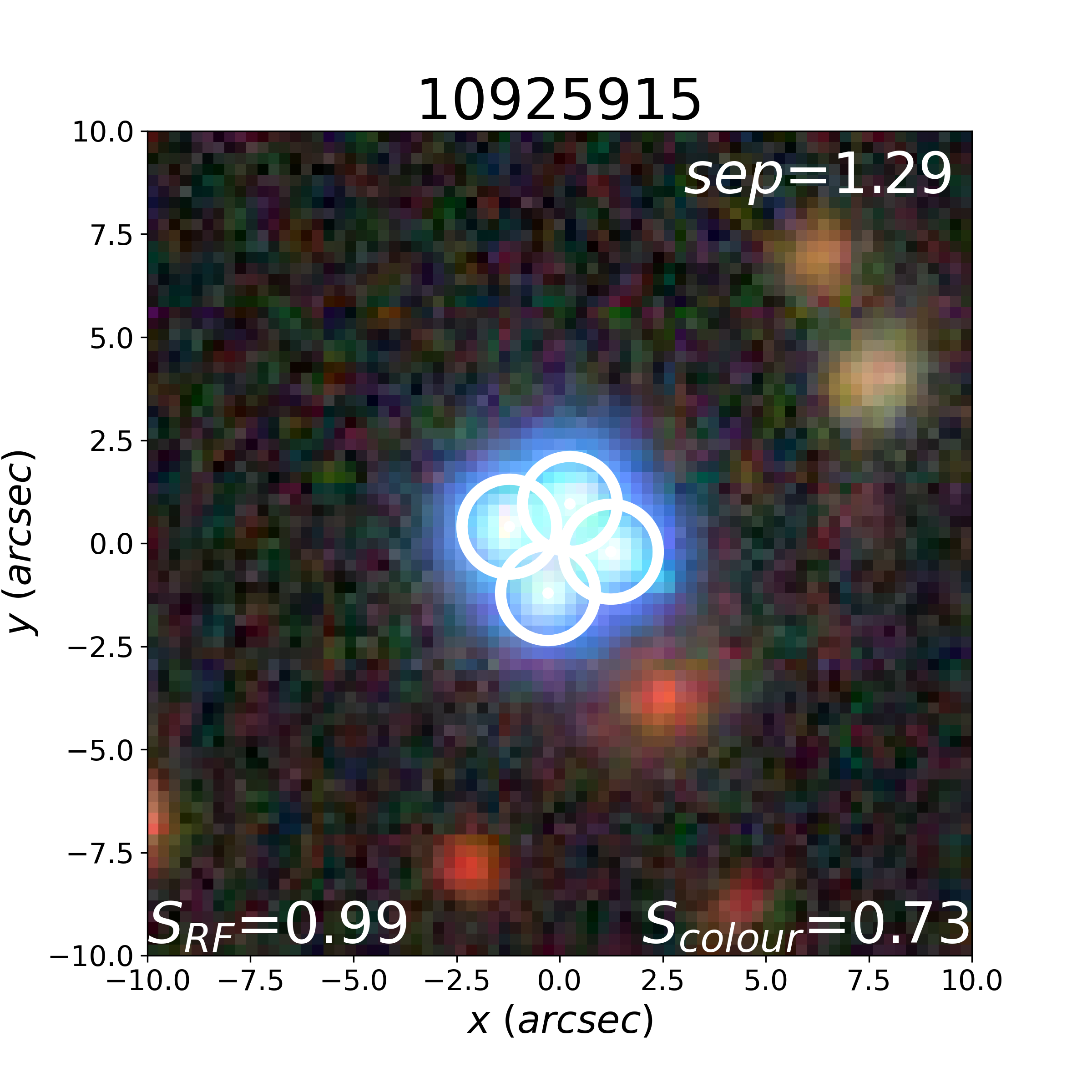}
        \centerline{\textbf{Example systems in both H22 and D22}}
        \includegraphics[scale=0.29]{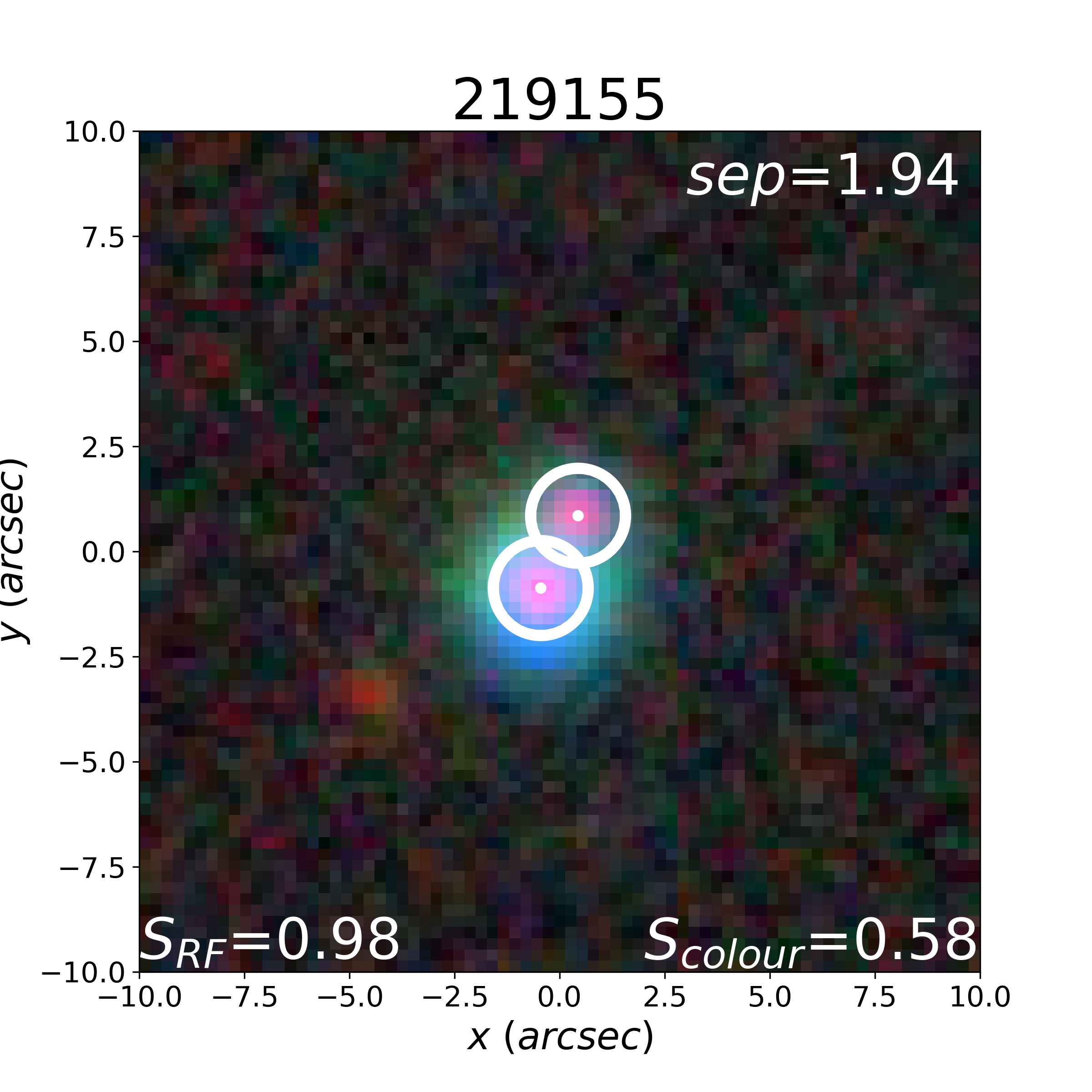}
        \includegraphics[scale=0.29]{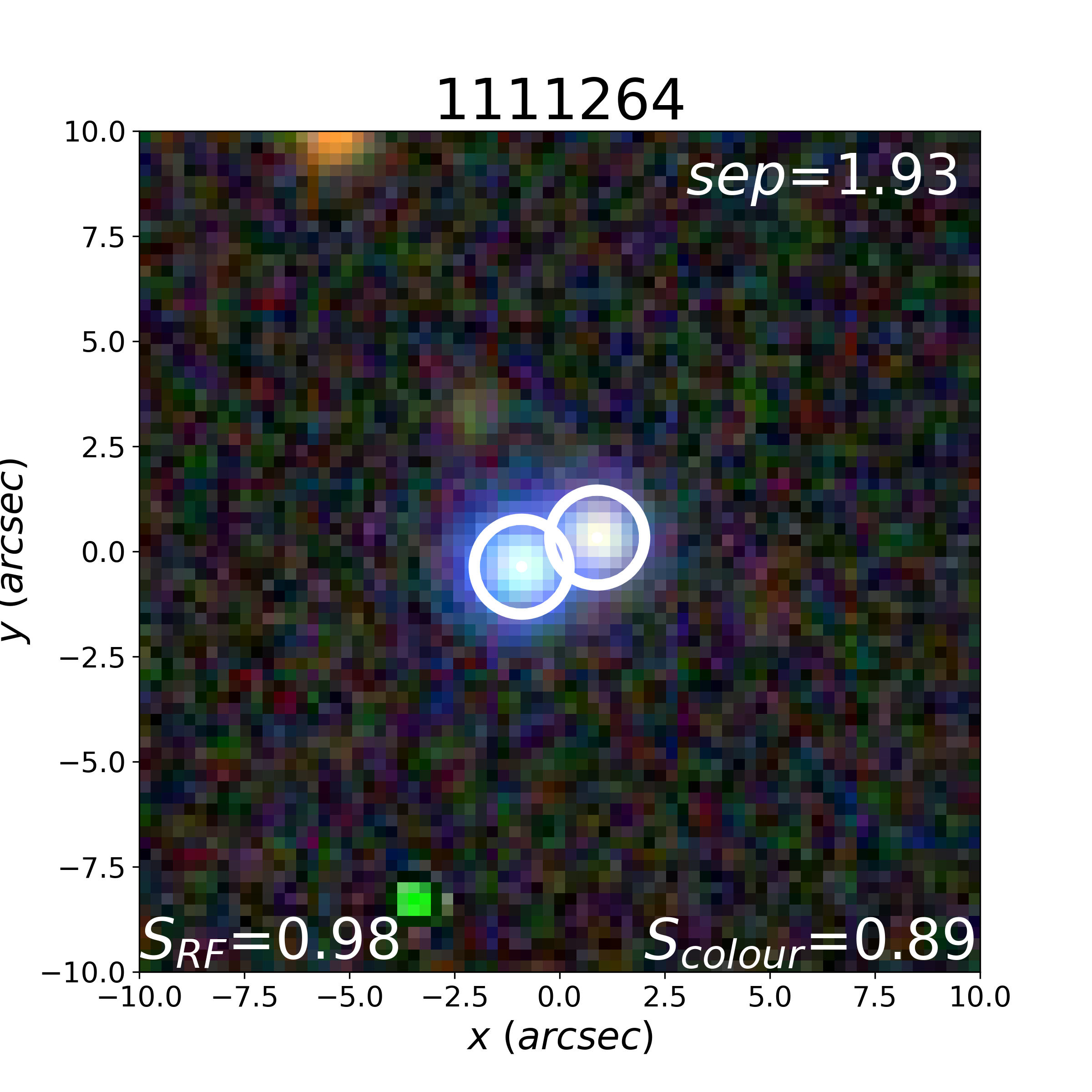}
        \includegraphics[scale=0.29]{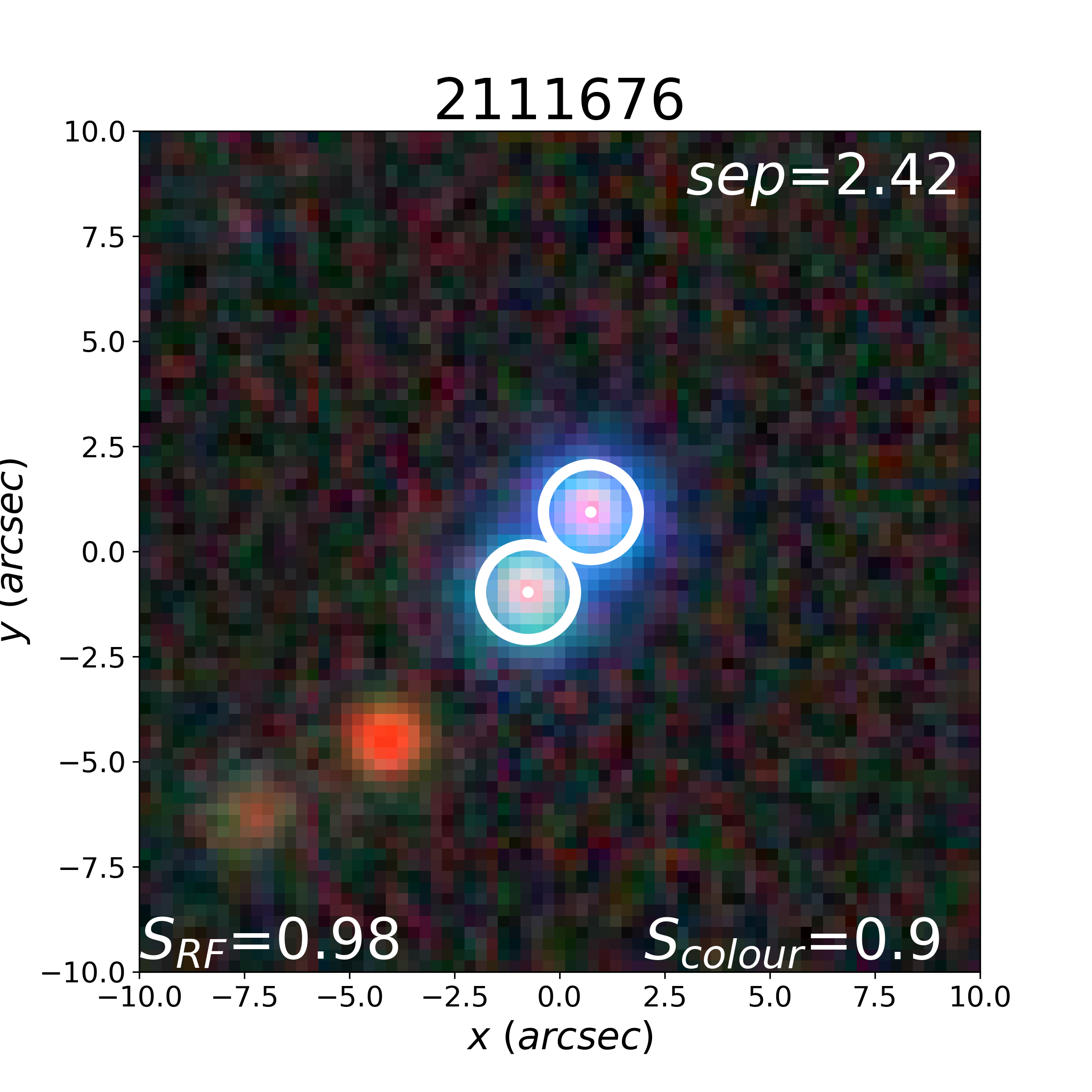}
        \centerline{\textbf{Example systems of new candidates not found in either of D22 or RLQ}}
        \includegraphics[scale=0.29]{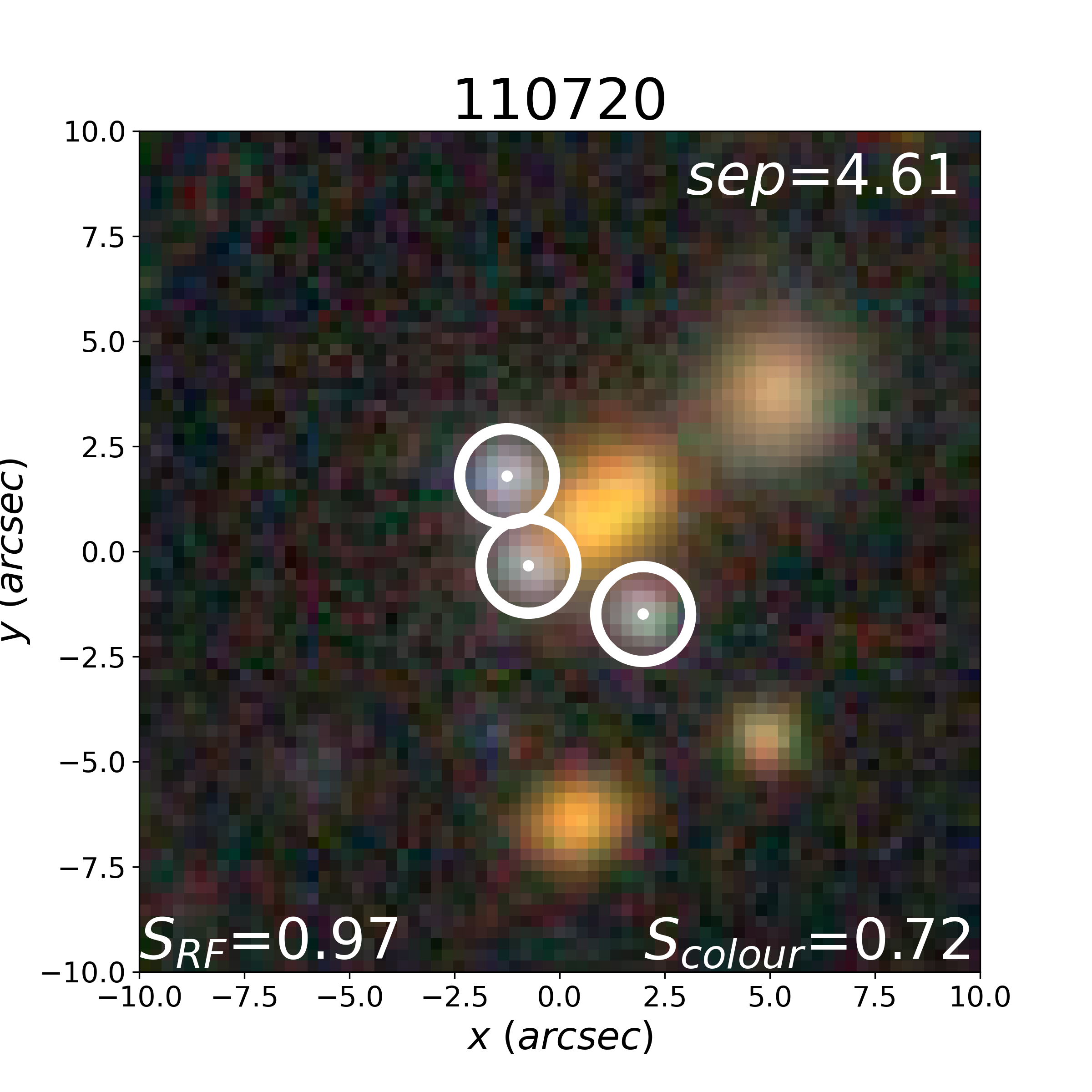}
        \includegraphics[scale=0.29]{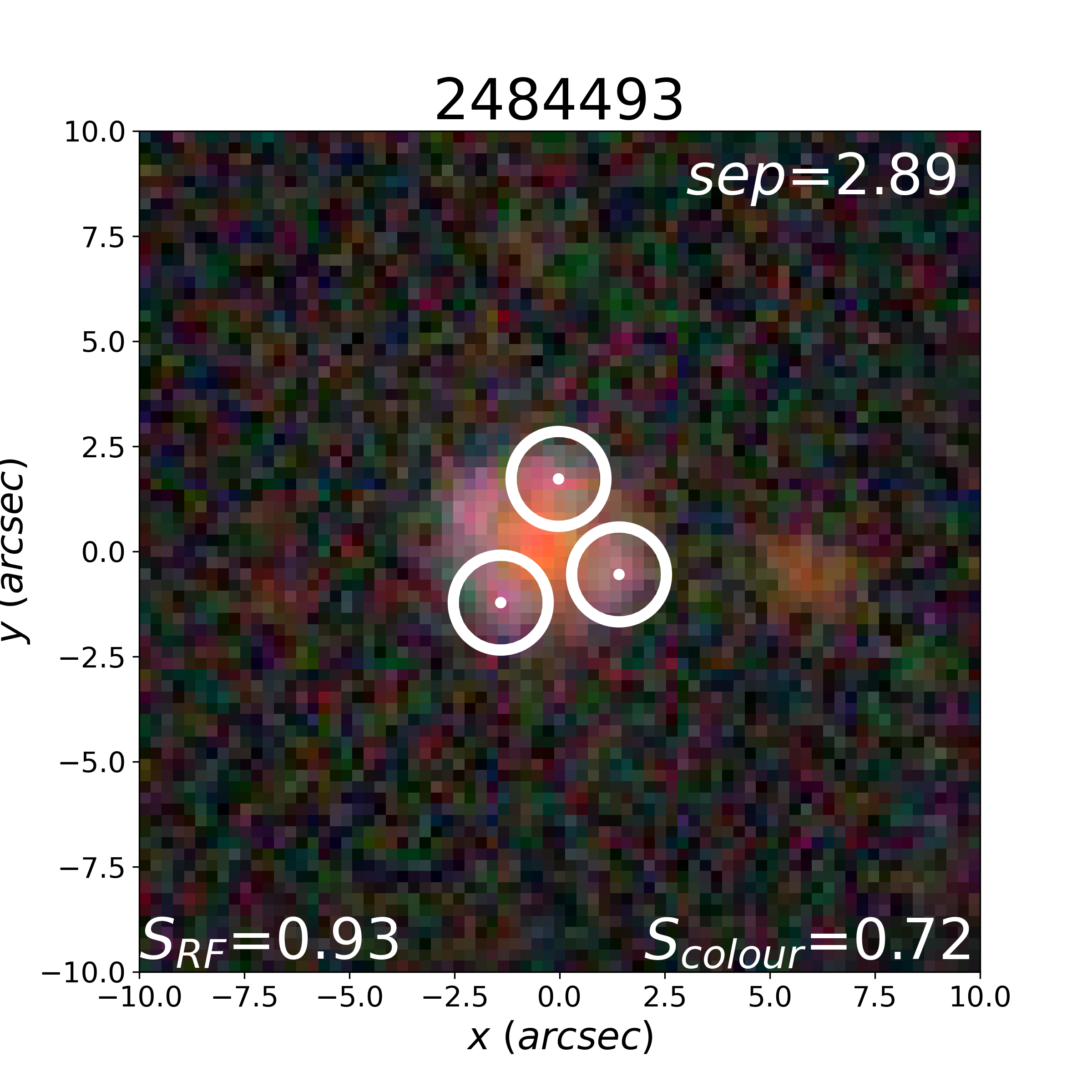}
        \includegraphics[scale=0.29]{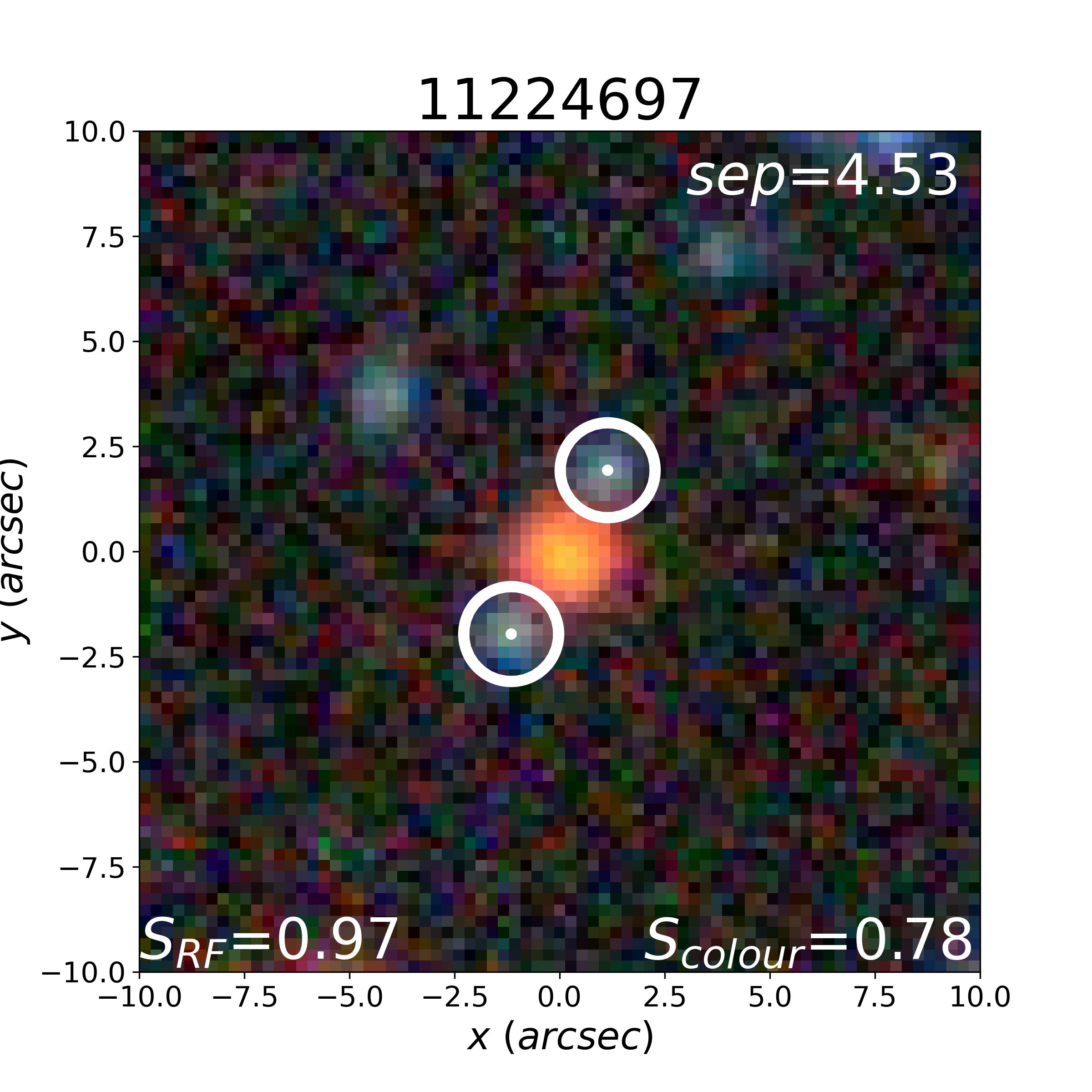}
        \caption{Examples of Grade A candidates in H22. The top row shows candidates also found in RLQ, the middle row shows candidates also found in the candidate catalogue of D22 and the bottom row shows new systems neither contained in RLQ nor D22. Each image is titled with the internal ID (linking the system to QGC) and gives the values of $S_{RF}$, $S_{colour}$ and image separation.}
        \label{fig:cands}
    \end{figure*}

    Fig.\ref{fig:cands} displays some example Grade-A systems. We show some examples of systems also found by D22, some confirmed systems in RLQ and finally some examples of new systems not found in D22 or RLQ. 
    
\begin{figure*}
        \centering
        \includegraphics[scale=0.45]{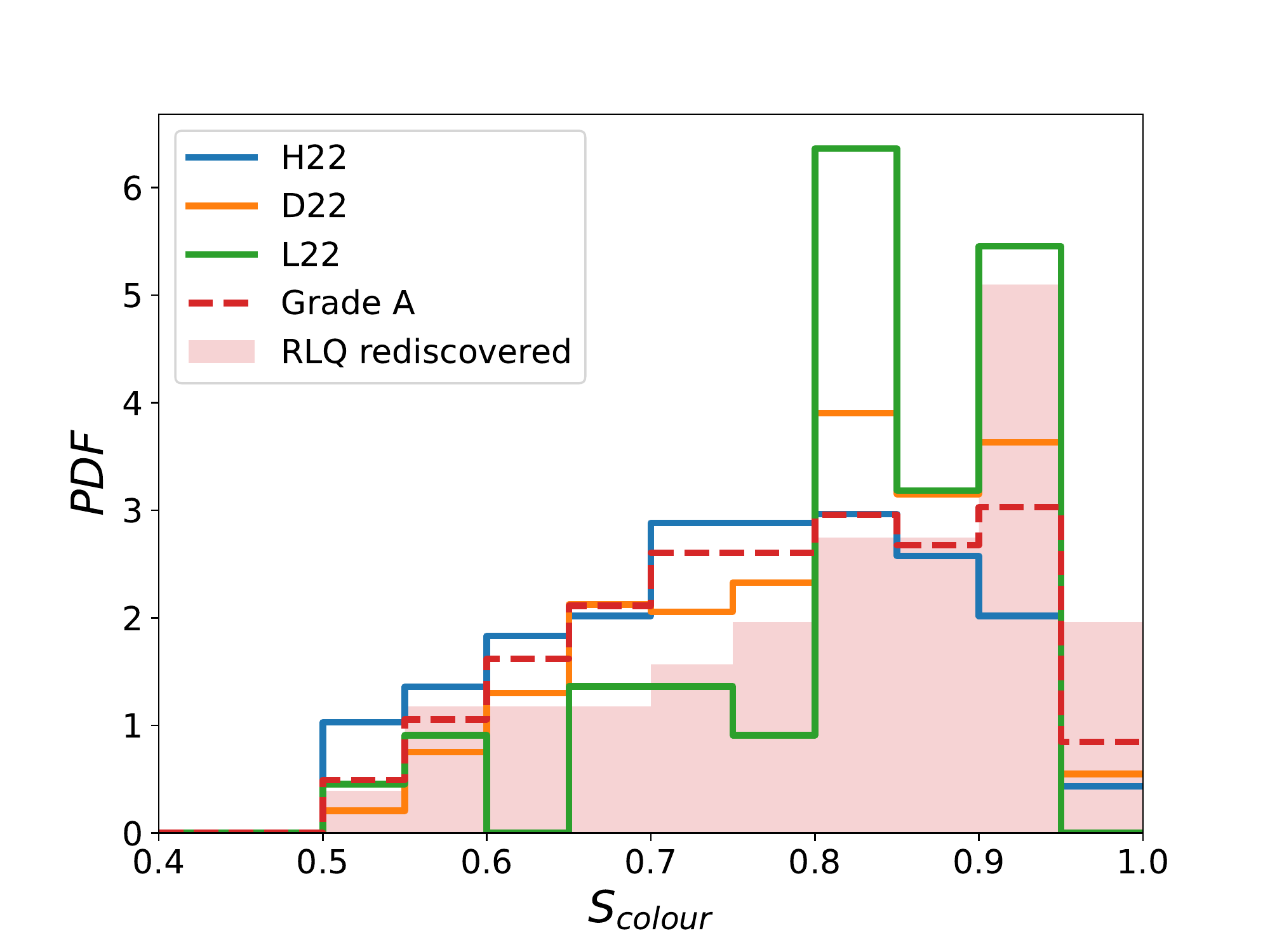}
        \includegraphics[scale=0.45]{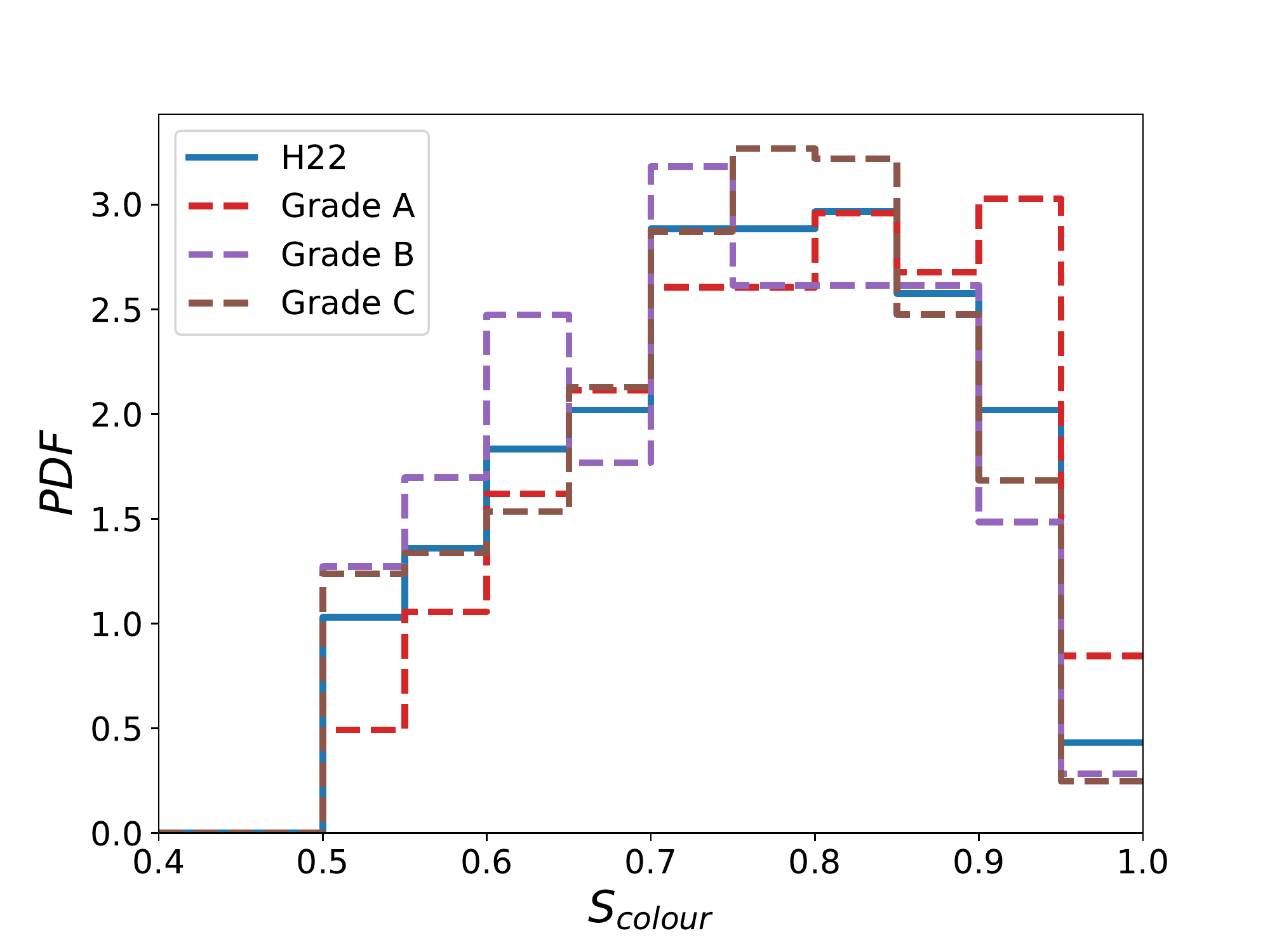}
        \includegraphics[scale=0.45]{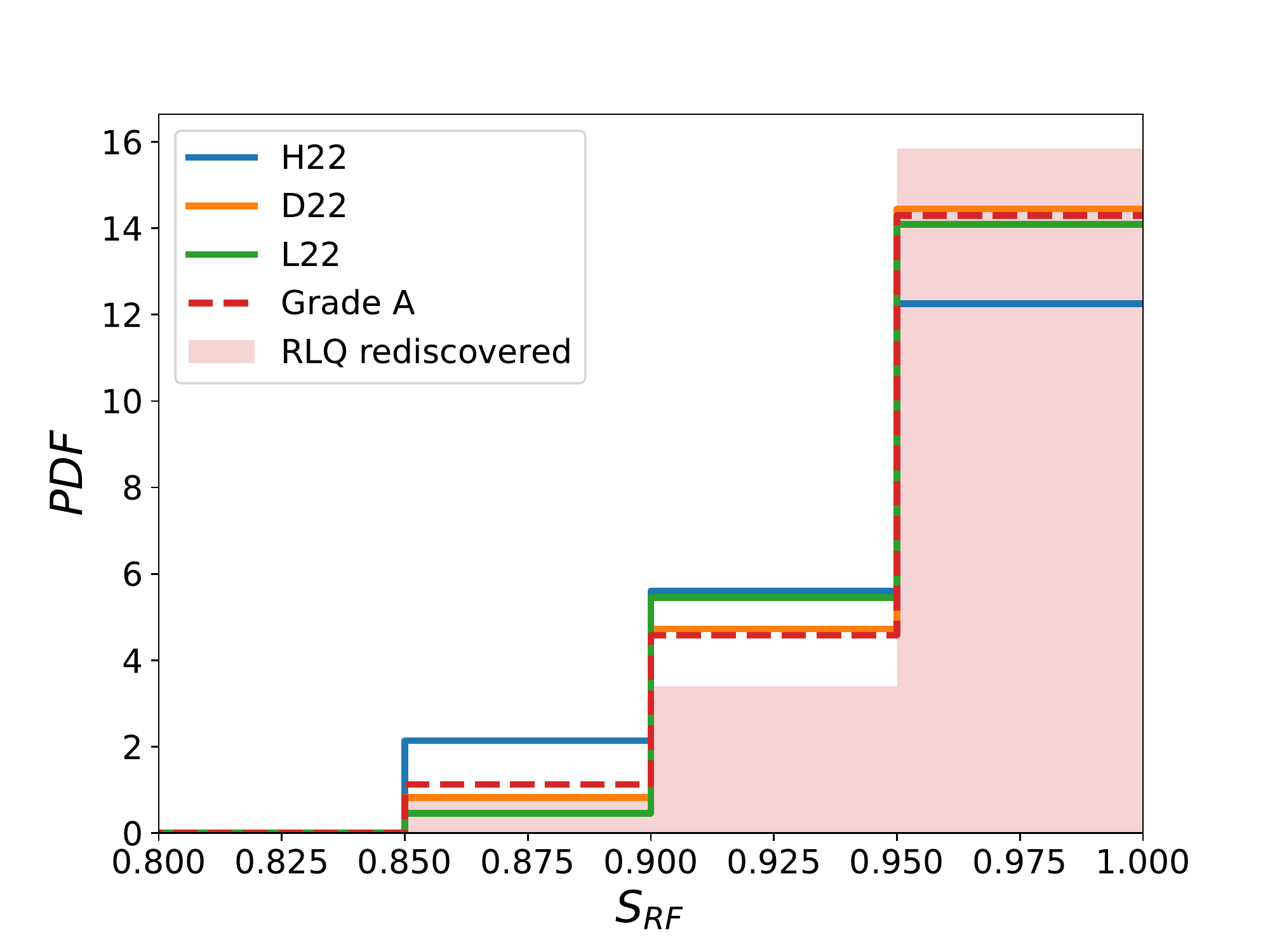}
        \includegraphics[scale=0.45]{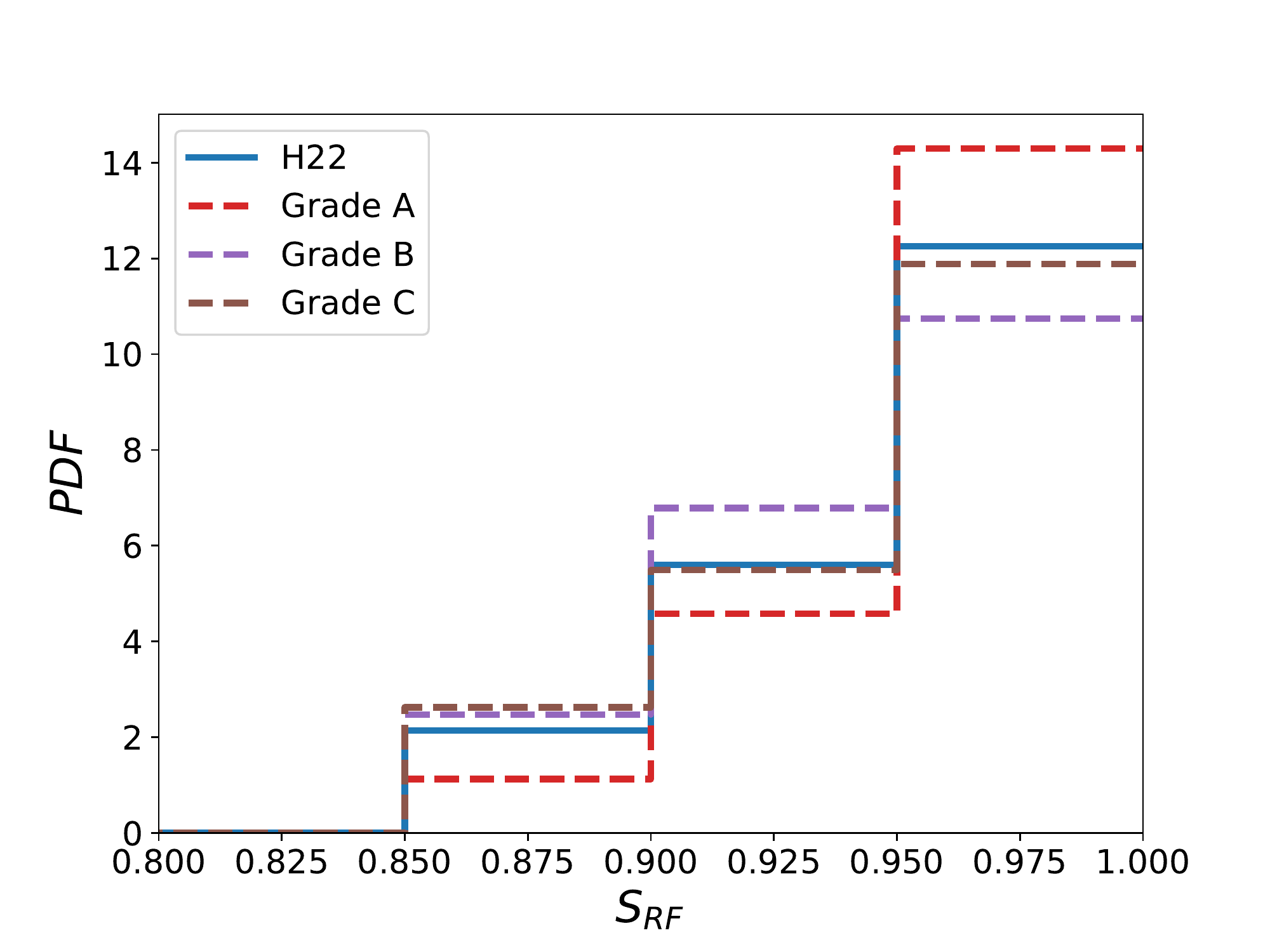}
        \caption{The distributions of $S_{colour}$ (top two panels) and $S_{RF}$ (bottom two panels) comparing  our new candidate systems (H22) with those of D22, L22, and RLQ (left panels) and splitting our new candidates by awarded grade (right panels).}
     \label{fig:H22vsD22vsL22_srf_sc}
    \end{figure*}

    In Fig.\,\ref{fig:H22vsD22vsL22_srf_sc}, we compare the distributions of $S_{colour}$ and $S_{RF}$ in H22 with those of D22, L22 and the `rediscovered' systems in RLQ that are found in H22. These are all statistically similar. The figure also shows the distributions of $S_{colour}$ and $S_{RF}$ split by grade in H22. As is expected, the Grade-A candidates match RLQ best, while Grade-Bs and Cs have similar distributions and differ more from RLQ.

    The redshift distribution and $g$-band magnitude distribution are shown in Fig.\,\ref{fig:H22vsD22vsL22_z} and Fig.\,\ref{fig:H22vsD22vsL22_mag} respectively and compared with OM10, L22, and D22. The source redshifts are acquired from the QSO catalogue of eBOSS DR16 \citep[][DR16Q hereafter]{Lyke2020}. 195 out of 971 entries of H22 have at least one match in DR16Q, and thus the corresponding source redshifts are available. In source redshift space, our samples are mostly distributed within $[0,3.5]$. Our method tends to select the samples at lower redshifts ($<2.0$), missing high-redshift lensed QSOs because of the depth limit of DESI-LS and WISE. The $g$-band magnitude distribution reveals that the luminosity distribution of the QSOs in H22 deviates from the power-law trend given by OM10 around $g \simeq 20.5$, indicating again that our method becomes insufficient beyond the depth limits of the observations (mainly limited by WISE data).
    
    Moreover, among the 195 candidates with redshift detections, 158 candidates have one redshift detection, and 37 (23 As + 5 Bs + 9 Cs) systems have two. The redshift overlaps between DR16Q and H22 give us some clues about the false positive rates of H22. Among 23 Grade-As, 21 have similar redshifts (difference $< 0.02$), but for 5 Grade-Bs, the number is 3 and for 9 Grade-Cs, the number is 5. Assuming that the candidates with two similar redshift detections are more likely to be the true lensed QSOs, this suggests that the false positive rate is increasing from A to B to C, although the trend is of course subject to a relatively large Poisson error.
    
    Distributions of image separations plotted in Fig.\,\ref{fig:H22vsD22vsL22_sep} show that there are significant differences between grades. The image separation distribution of Grade-As peaks at $\sim$ 0.8 arcsec, while those of Grade-Bs and Cs peak at $\sim$ 2.2 arcsec. The dissimilarity is due to the criteria applied during human visual inspection (Sec.\,\ref{sec:vs}) which assigns a lower grade to larger image separations. The distribution of Grade A image separations agrees with OM10, but those of Grade-B and C do not. Contamination by stars is more likely at larger image separations and therefore the false positive rate in the Grade A lenses is expected to be lower.
    
    \begin{table*}[]
    \centering
    \begin{tabular}{|cc|}
    \hline
    \multicolumn{2}{|c|}{All candidates}                                                                                                                  \\ \hline
    \multicolumn{2}{|c|}{\textit{This catalogue contains 971 candidates identified in this work}}                                                    \\ \hline
    \multicolumn{1}{|c|}{Columns}                 & Description                                                                                           \\ \hline
    \multicolumn{1}{|c|}{internal\_ID}            & Internal ID that links this system to QGC                                                           \\ \hline
    \multicolumn{1}{|c|}{RA}                      & Right Ascension (J2000)                                                                               \\ \hline
    \multicolumn{1}{|c|}{Dec}                     & Declination (J2000)                                                                                   \\ \hline
    \multicolumn{1}{|c|}{S\_RF}                   & One of the scores defined in Sec.\,\ref{sec:select_groups_bycolor}                         \\ \hline
    \multicolumn{1}{|c|}{S\_colour}               & One of the scores defined in Sec.\,\ref{sec:select_groups_bycolor}                          \\ \hline
    \multicolumn{1}{|c|}{Sep\_max}                & Maximum image separation in arcseconds                                                                \\ \hline
    \multicolumn{1}{|c|}{Sep\_min}                & Minimum image separation in arcseconds                                                                \\ \hline
    \multicolumn{1}{|c|}{Sep\_mean}               & Mean image separation in arcseconds                                                                   \\ \hline
    \multicolumn{1}{|c|}{Grade}                   & Grades of lensed QSO candidates (A, B, or C)                                                       \\ \hline
    \multicolumn{1}{|c|}{Grade\_by\_LR}           & Grades of lensed QSO candidates (A, B, or C) that are given independently by the fourth author     \\ \hline
    \multicolumn{1}{|c|}{NUM\_of\_IMGs}           & Number of images                                                                                      \\ \hline
     \multicolumn{1}{|c|}{in\_RLQ}                 & Whether this candidate exists in RLQ                                                                  \\ \hline
    \multicolumn{1}{|c|}{in\_D22}                 & Whether this candidate exists in D22                                                                  \\ \hline
    \multicolumn{1}{|c|}{in\_L22}                 & Whether this candidate exists in L22                                                                  \\ \hline
    \multicolumn{1}{|c|}{Grade\_by\_D22}          & Grades of lensed QSO candidates (A, B, or C) given in D22                                          \\ \hline
    \multicolumn{1}{|c|}{Classification\_by\_L22} & The classification in L22 (lens, NIQ, projected QSOs, or QSO pairs)                                   \\ \hline
    \multicolumn{1}{|c|}{$q$}                     & Axis ratio of SIE profile given by lens modelling                                                     \\ \hline
    \multicolumn{1}{|c|}{phi}                     & Position angle of SIE profile (+x-axis is zero, counterclockwise is positive) given by lens modelling \\ \hline
    \multicolumn{1}{|c|}{rein}                    & Einstein radius  given by lens modelling                                                              \\ \hline
    \multicolumn{1}{|c|}{minimised $\chi^2$}      & The $\chi^2$ that calculated by the best-fitted lens model and observation data                       \\ \hline
    \multicolumn{1}{|c|}{BIC}      & The Bayesian information criterion that is defined in Sec.\,\ref{sec:lens_modelling}.       \\ \hline
    \multicolumn{1}{|c|}{Note}                    & Comments on this lensed QSO candidate                                                              \\ \hline
    \end{tabular}
    \caption{Descriptions of the properties provided for each lensed QSO system in our final catalogue H22. The catalogue contains new candidates and includes rediscovered confirmed systems. The full catalogue is available online (see footnote 10).}

    \label{tab:all_cand}
    \end{table*}
    \begin{table*}[]
        \centering
        \begin{tabular}{|cc|}
        \hline
        \multicolumn{2}{|c|}{\textbf{New candidates in detail}}                                                                                   \\ \hline
        \multicolumn{2}{|c|}{\textit{This catalogue describes every images (1977 in total) of each candidate in H22}} \\ \hline
        \multicolumn{1}{|c|}{Columns}                             & Description                                                                   \\ \hline
        \multicolumn{1}{|c|}{internal\_ID}                           & Internal ID that links this system to QGC                                  \\ \hline
        \multicolumn{1}{|c|}{RA}                                  & Right Ascension (J2000) of the centre of this image                           \\ \hline
        \multicolumn{1}{|c|}{Dec}                                 & Declination (J2000) of the centre of this image                               \\ \hline
        \multicolumn{1}{|c|}{$g$}                                 & $g$-band magnitude of this image                                              \\ \hline
        \multicolumn{1}{|c|}{$r$}                                 & $r$-band magnitude of this image                                              \\ \hline
        \multicolumn{1}{|c|}{$z$}                                 & $z$-band magnitude of this image                                              \\ \hline
        \multicolumn{1}{|c|}{$W1$}                                & $W1$-band magnitude of this image                                             \\ \hline
        \multicolumn{1}{|c|}{$W2$}                                & $W2$-band magnitude of this image                                             \\ \hline
        \multicolumn{1}{|c|}{redshift}                            & The matched redshift from eBOSS DR16 (if available)                                \\ \hline
        \multicolumn{1}{|c|}{Grade}                               & The grades (A,B,C) of the candidate system that this image belongs to.      \\ \hline
        \end{tabular}
        \caption{Descriptions of the properties given for each image of each candidate lensed QSO in the detailed catalogue of H22-details. The full catalogue is online available (see footnote 11).}
    \label{tab:cand_details}
    \end{table*}

    \subsection{New lensed QSOs candidates}
    \label{sec:new_to_RLQ}
    
    Our catalogue of candidate multiply imaged lensed QSOs contains 620 new systems not contained in the RLQ catalogue, L22 or D22. In this section, we discuss the key differences between our new catalogue, H22, and the RLQ catalogue and those of D22 and L22.
    
    \subsubsection{Compared to RLQ} 
    
    Compared to RLQ, we have identified 918 extra candidates out of 971 and labelled them as H22-new-RLQ hereafter (see Fig.\,\ref{fig:cands}). Note that some of these 918 candidates are also found in D22 and L22; we discuss these overlaps in sections \ref{sec:subsub2D22} and \ref{sec:subsub2L22}.

    Comparisons of the distributions of $S_{colour}$ and $S_{RF}$ between H22 and RLQ  are shown in Fig.\,\ref{fig:H22vsD22vsL22_srf_sc}. At the upper end of $S_{colour}$, H22 shows a more significant difference to RLQ, which indicates that there are items with low $S_{colour}$ in H22. This is likely due to false positives which spectroscopic follow-up would reject.  In contrast, the $S_{RF}$ statistic of RLQ shows a much more similar distribution to our catalogue, implying that $S_{RF}$ is a more distinguishing criterion in the identification process.
    
    The redshift and magnitude distributions of strongly lensed images in the candidate systems are displayed in Fig.\,\ref{fig:H22vsD22vsL22_z} and Fig.\,\ref{fig:H22vsD22vsL22_mag}. We find that H22 covers broader redshift and magnitude ranges than RLQ, for example, H22's redshift range beyond 1.8 and magnitude range beyond $g\simeq 19$. Comparing the max image separation distributions between H22 and RLQ (Fig.\,\ref{fig:H22vsD22vsL22_sep}), we find that H22 has a wider distribution than RLQ. Notably, the lower end is filled with Grade-A candidates (whose distribution matches the theoretical predictions of OM10), while Grade-B and Grade-C candidates dominate the upper end (whose distributions differ from OM10). 

    The majority of systems in H22-new-RLQ are pairs. A few possible quads (110720, 2484493, 11419327, 11125158) exist. Thus, the quads-to-pair ratio ($\sim 1/30$) is unexpectedly lower than the theoretical prediction of $\sim 1/6$ given by OM10. Multiple causes may lead to such an issue. One possibility is that the quadruple systems are more prone to the effects of PSF smearing than duals; when the max image separations are comparable to the PSF size, it is harder to distinguish quads than duals which likely causes quads to be under-respresented in the QCC (see Fig.\,\ref{fig:demo_of_miss} and the corresponding discussion). It is likely that  adopting image-based deep learning approaches similar to those developed for searching galaxy-galaxy strong lensing systems \citep[e.g.,][]{Petrillo2017,Lanusse2018,He2020,Li2021,Huang2021,Rojas2022} would help retrieve these missing quads. Another possibility is that a low quads-to-duals ratio could be caused by the dual candidates having a higher false positive rate than the quad candidates. In principle, the probability of the emergence of two-image systems of non-lenses is significantly higher than that of four-image systems, especially when a requirement of visual inspection is that the quads must have typical lensed-image configurations. Hence, our candidate catalogue is unsuitable for estimating the double-to-quad ratio but reveals the issues that need to be improved.
    
    \subsubsection{Compared to D22} 
    \label{sec:subsub2D22}
    
\begin{figure}
        \centering
        \includegraphics[scale=0.45]{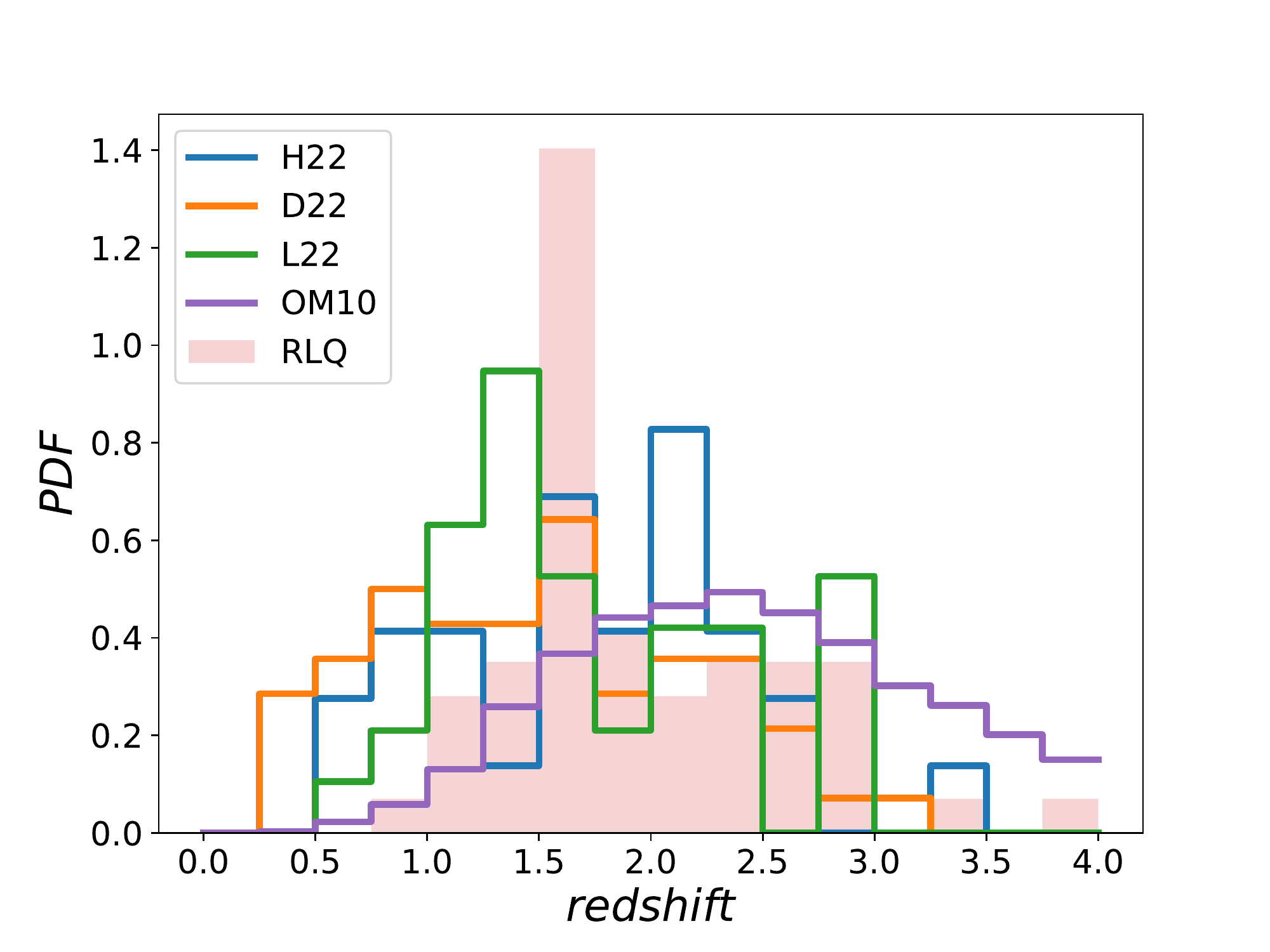}
        \caption{The comparison of redshift distributions of H22, D22, L22, and OM10.  We note that the redshifts of L22 are from \cite{lemon2022}, while those of H22 and D22 are from SDSS eBOSS DR16.}
        \label{fig:H22vsD22vsL22_z}
    \end{figure}

    D22\footnote{\url{https://sites.google.com/usfca.edu/neuralens/publications/lensed-qso-candidates-dawes-2022}} is another catalogue of lensed QSO candidates extracted from DESI Legacy Imaging Surveys, created by \cite{Dawes2022}. This study uses an independent approach to ours and contains 436 candidates. To understand the differences between the selection effects of H22 and D22, we compare distributions of $S_{colour}$, $S_{RF}$ (Fig.\,\ref{fig:H22vsD22vsL22_srf_sc}), source redshift (Fig.\,\ref{fig:H22vsD22vsL22_z}), $g$-band magnitude (Fig.\,\ref{fig:H22vsD22vsL22_mag}), and max separation (Fig.\,\ref{fig:H22vsD22vsL22_sep}). 
    
    There are 292 matching systems between H22 and D22 (which can be extracted by requiring `in\_D22=True' in H22). According to our grading, 133 are Grade-A, 62 are Grade-B, and 97 are Grade-C. However, according to D22's grading, 76, 83, and 133 out of 292 matches are Grade-A, B, and C respectively. We include the grade labels of D22 in H22. Given that the grades are assigned by human inspectors independently, the above difference are a reflection of the subjective nature of this process.
    
    The 679 systems in H22 not found in D22 are labelled as H22-new-D22. These contain 151 Grade-As, 221 Grade-Bs, and 307 Grade-Cs. However, there are 144 candidates proposed in D22 but missed by H22. Specifically, 43 systems are rejected by our selection by $S_{colour}$ and $S_{RF}$; $S_{colour}$ causes 38 rejections while $S_{RF}$ causes 5. The other 100 systems are not present in the QGC from the start. To summarise, 100 out of 144 mismatches are caused by different parent samples, 43 are caused by different selection methodologies and only one is caused by human inspection.
    
    The comparisons of the distributions of $S_{colour}$ and $S_{RF}$ between H22 and D22 in are shown in Fig.\,\ref{fig:H22vsD22vsL22_srf_sc}. We find that the distributions of D22 are closer to RLQ than H22. If we select only the Grade-A systems from H22, the distributions become similar to D22, suggesting that D22 has a slightly higher purity than H22. Fig.\,\ref{fig:H22vsD22vsL22_z} and \ref{fig:H22vsD22vsL22_mag} demonstrate that H22 is deeper than D22 in both magnitude and redshift, mostly due to the different parent samples adopted in the two works. D22 used the `DESI QSO Sample \citep{Yeche2020}' which has a hard magnitude cut of $r=22.7$ for the QSO detections. In contrast, QCC does not have any cuts in magnitude. To provide some quantification of the effect of this, there are $\sim 52\%$ candidates in QCC that are fainter than $r=22.7$. Regarding the distributions of image separation (Fig.\,\ref{fig:H22vsD22vsL22_sep}), D22's separation is smaller than H22's but larger than the Grade-A systems in H22. This indicates that H22-new-D22 (especially Grade-B and Grade-C systems in H22) covers multiply imaged lensed QSO systems with large image separations missed by D22 and thus potentially improves the completeness of the sample of multiple image QSOs systems from DESI Legacy Imaging Surveys.

    \subsubsection{Compared to L22} 
        \label{sec:subsub2L22}

   L22 is a catalogue that contains the spectroscopic follow-up results of 175 systems selected from a multiply imaged QSO candidate catalogue given by \cite{Lemon2019} based on Gaia DR2, which is an important update to GLQ. It takes advantage of the high astrometric precision of Gaia to discover many new lensed QSOs, including confirmed lensed QSOs, nearly identical QSOs (NIQ), projected QSOs, and QSO pairs. Following the definitions in \cite{lemon2022}, the two QSOs of an NIQ system have similar spectra but can not be confirmed as a strong lensing system because of the absence of a lens galaxy image. Nevertheless, the NIQs should be considered very promising candidates. `Projected QSO's occur when the redshifts of two QSOs are different, while `QSO pairs' are cases where the QSOs have similar redshifts but different spectra. For lensed QSO searching, QSO pairs and projected QSOs are contaminations. 

    There are 44 matches between L22 and H22. The 948 systems in H22 not found in L22 are labelled as H22-new-L22. In the 44 matches, 17 systems are labelled as lenses in L22, 20 are NIQs, 2 are projected QSOs, and 3 are QSO pairs. In other words, 5 systems are false positives, and 39 are confirmed lenses or promising candidates. The comparisons of the distributions of $S_{colour}$ and $S_{RF}$ between H22 and L22 are shown in Fig.\,\ref{fig:H22vsD22vsL22_srf_sc}. The figure shows that compared to H22 and D22, L22 has the most similar distributions to RLQ. This is unsurprising because most of the systems in L22 are spectroscopically confirmed lenses and NIQ. 
    
\begin{figure*}
        \centering
        \includegraphics[scale=0.45]{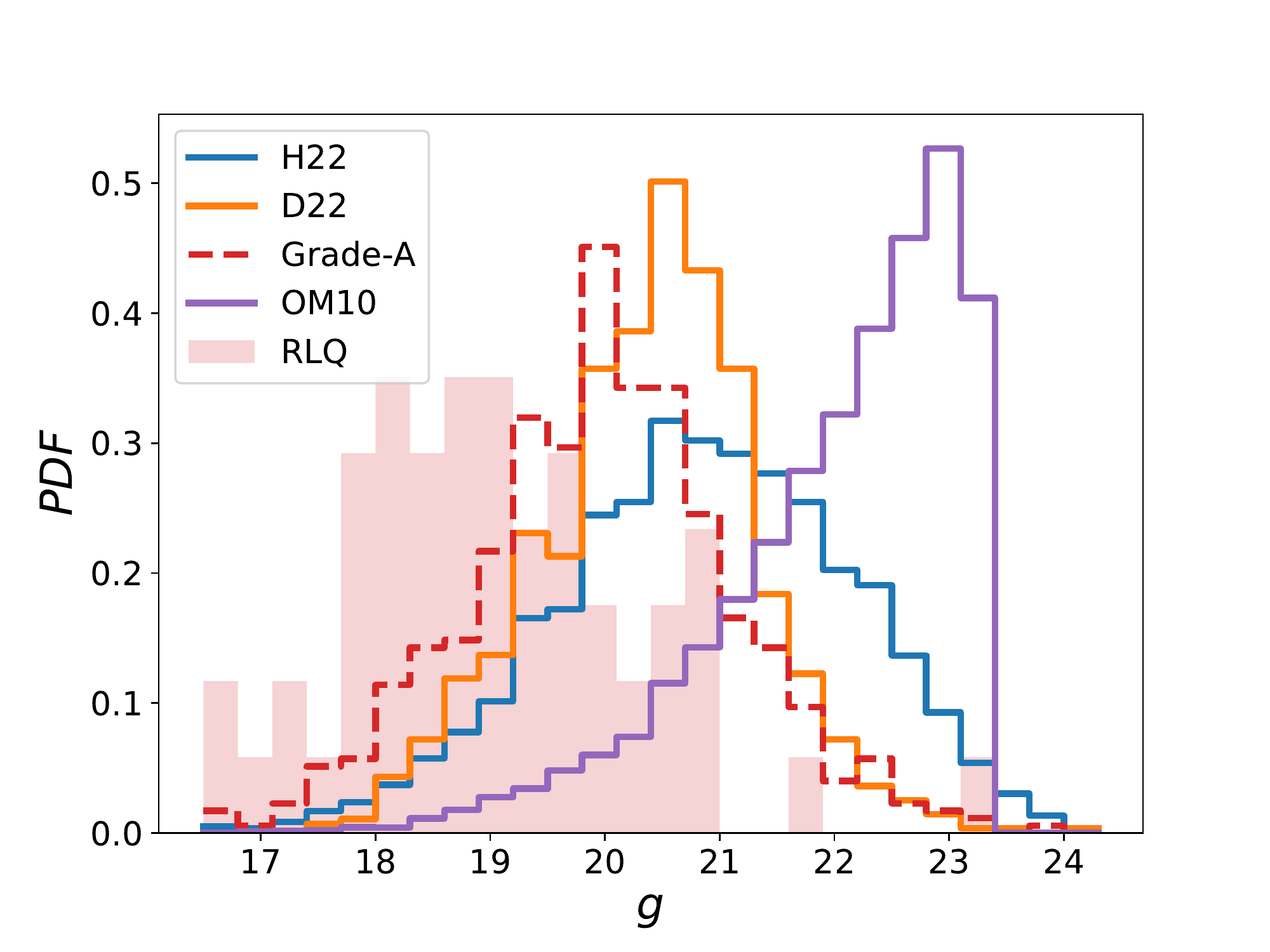}
        \includegraphics[scale=0.45]{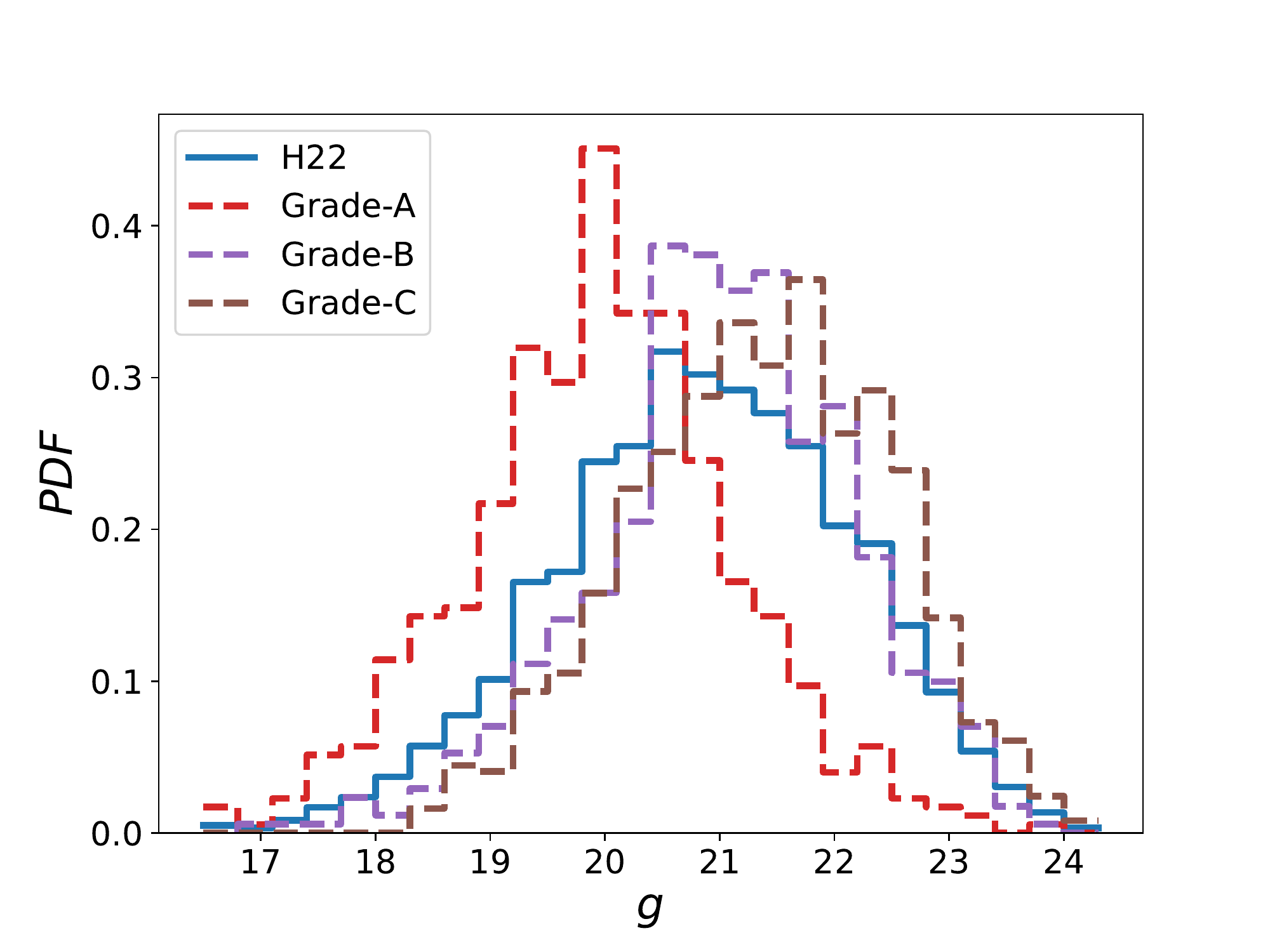}
        \caption{Left panel: Comparison of magnitude distributions of H22, D22, and OM10. Since L22 did not provide $g$-band magnitudes, it is not plotted here. Right panel: Distributions split by awarded grade in H22.}
        \label{fig:H22vsD22vsL22_mag}
    \end{figure*}

\begin{figure*}
        \centering
        \includegraphics[scale=0.45]{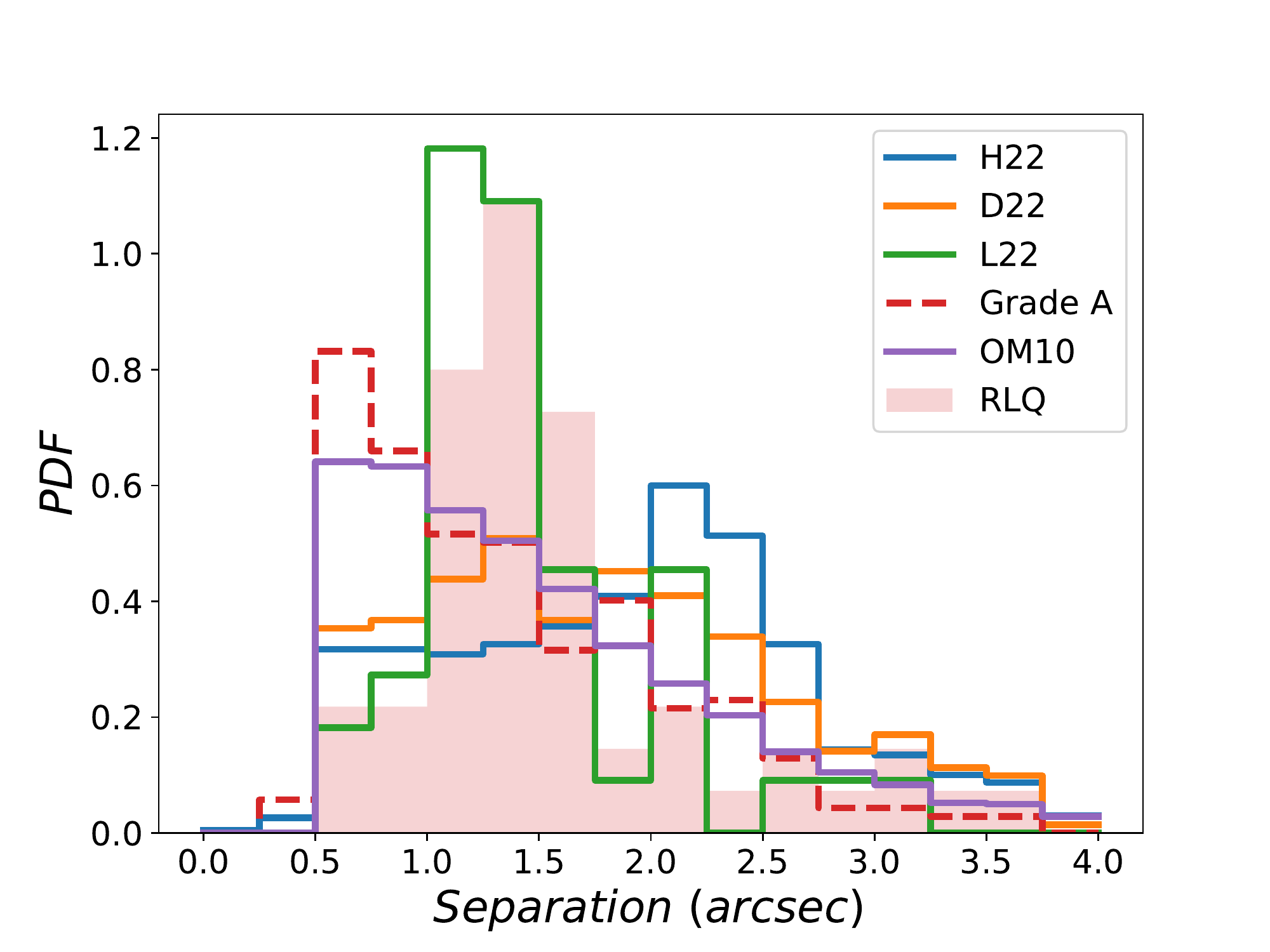}
        \includegraphics[scale=0.45]{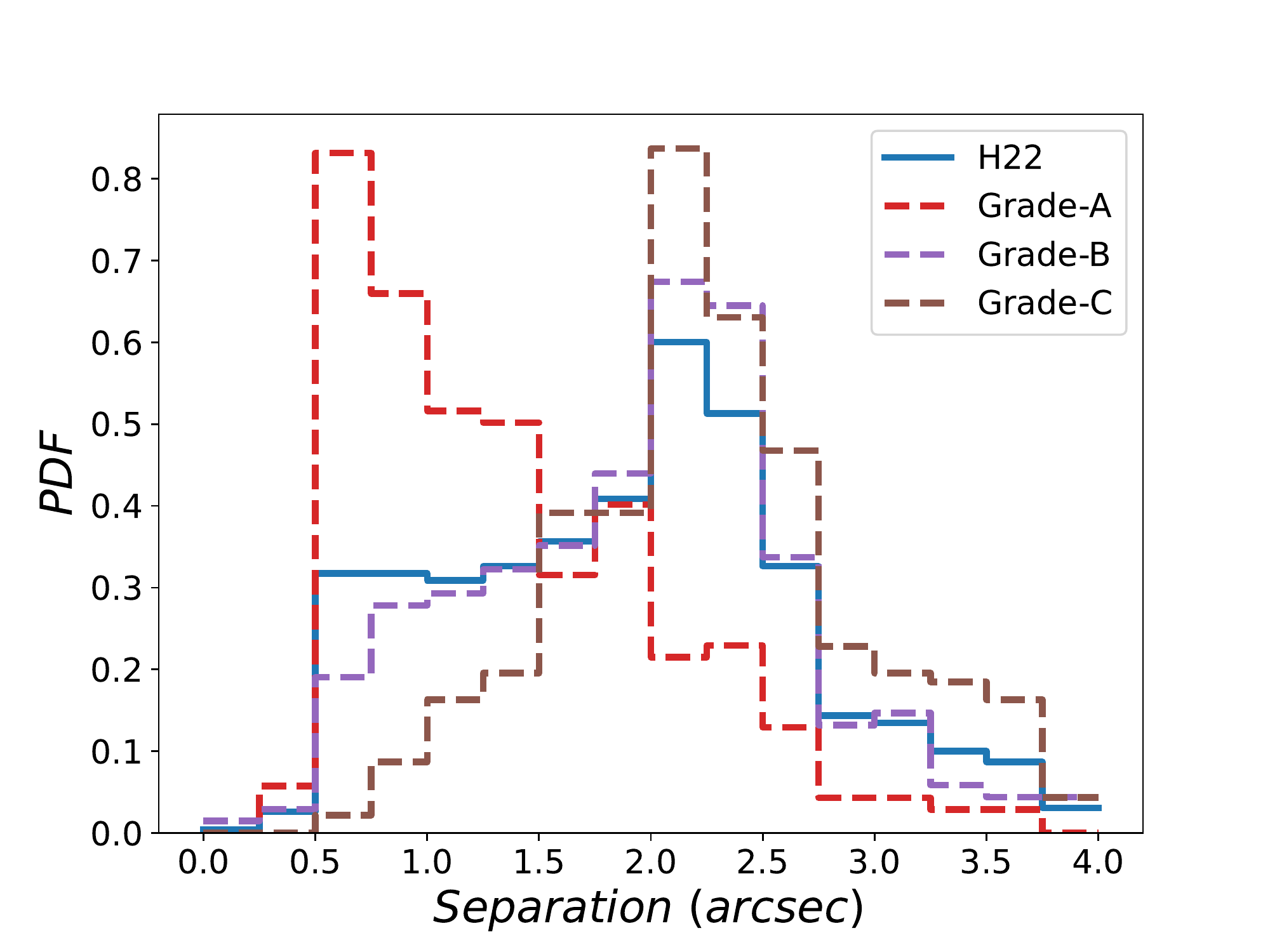}
        \caption{Comparison of max image separation of systems in H22, D22, L22, RLQ, and OM10. The panel on the right splits this distribution for H22 by awarded grade.}
        \label{fig:H22vsD22vsL22_sep}
    \end{figure*}

    Fig.\,\ref{fig:H22vsD22vsL22_mag} shows that L22 is shallower than H22 and D22 (i.e. lower mean redshift). This is due to their selection from shallower Gaia data \citep{GaiaCollaboration2018}. The image separations of L22 are similar to those of RLQ, which again is a reflection of the fact that L22 primarily consists of spectroscopically confirmed lensed QSO systems (Fig.\,\ref{fig:H22vsD22vsL22_sep}).


    \subsubsection{New candidates compared to RLQ, D22, and L22}

\begin{figure*}
        \centering
        \includegraphics[scale=0.45]{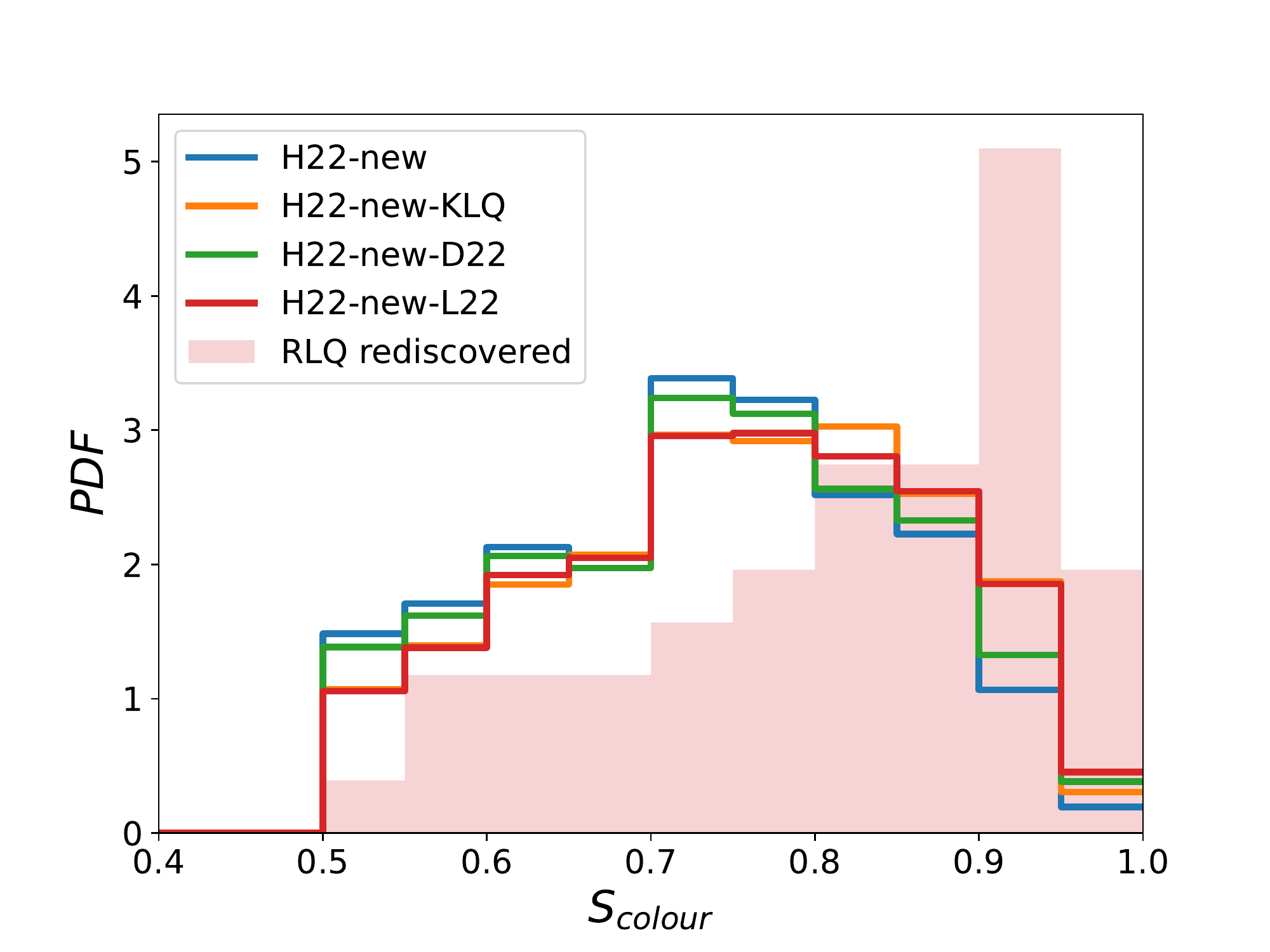}
        \includegraphics[scale=0.45]{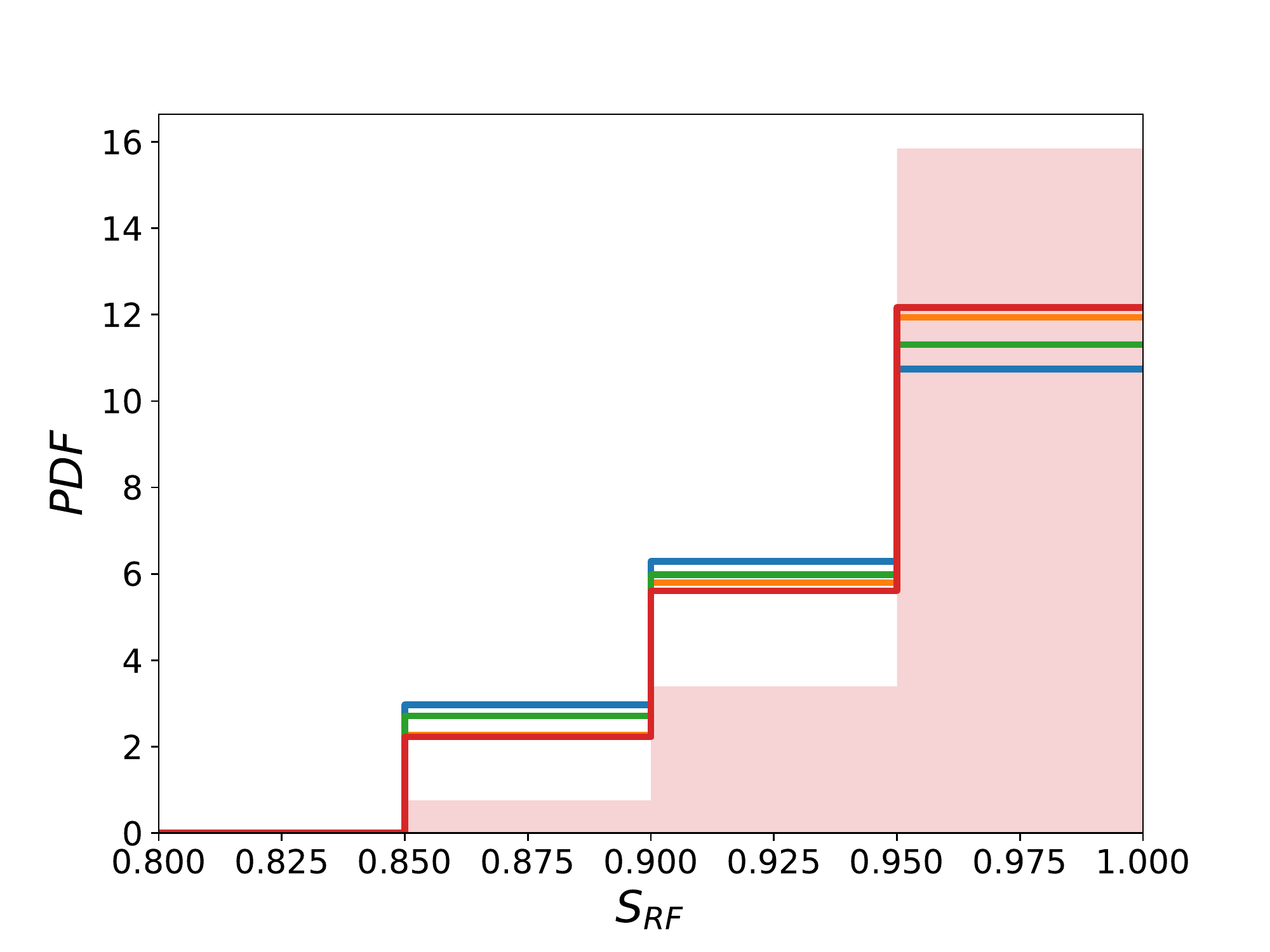}
        \includegraphics[scale=0.45]{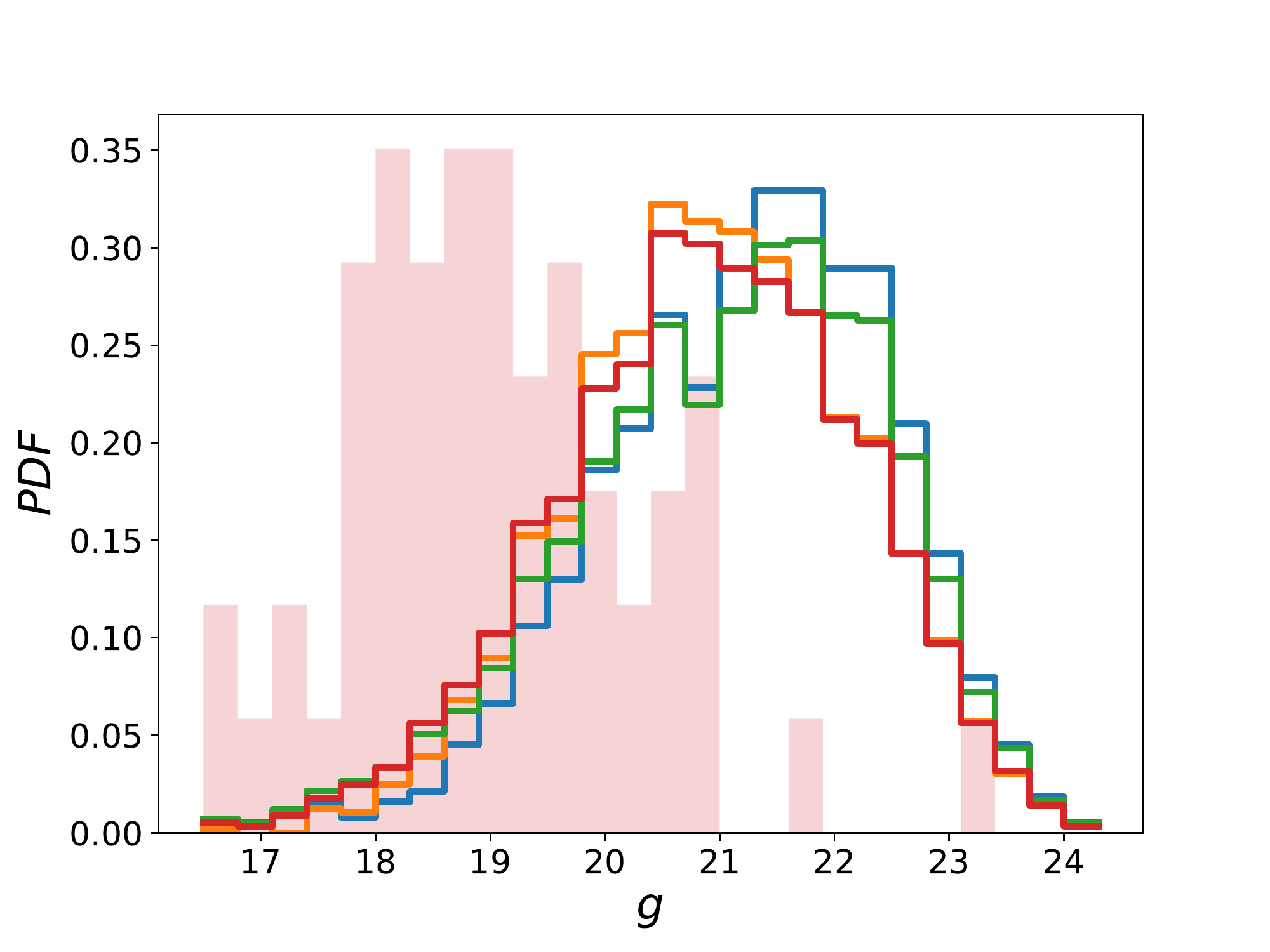}
        \includegraphics[scale=0.45]{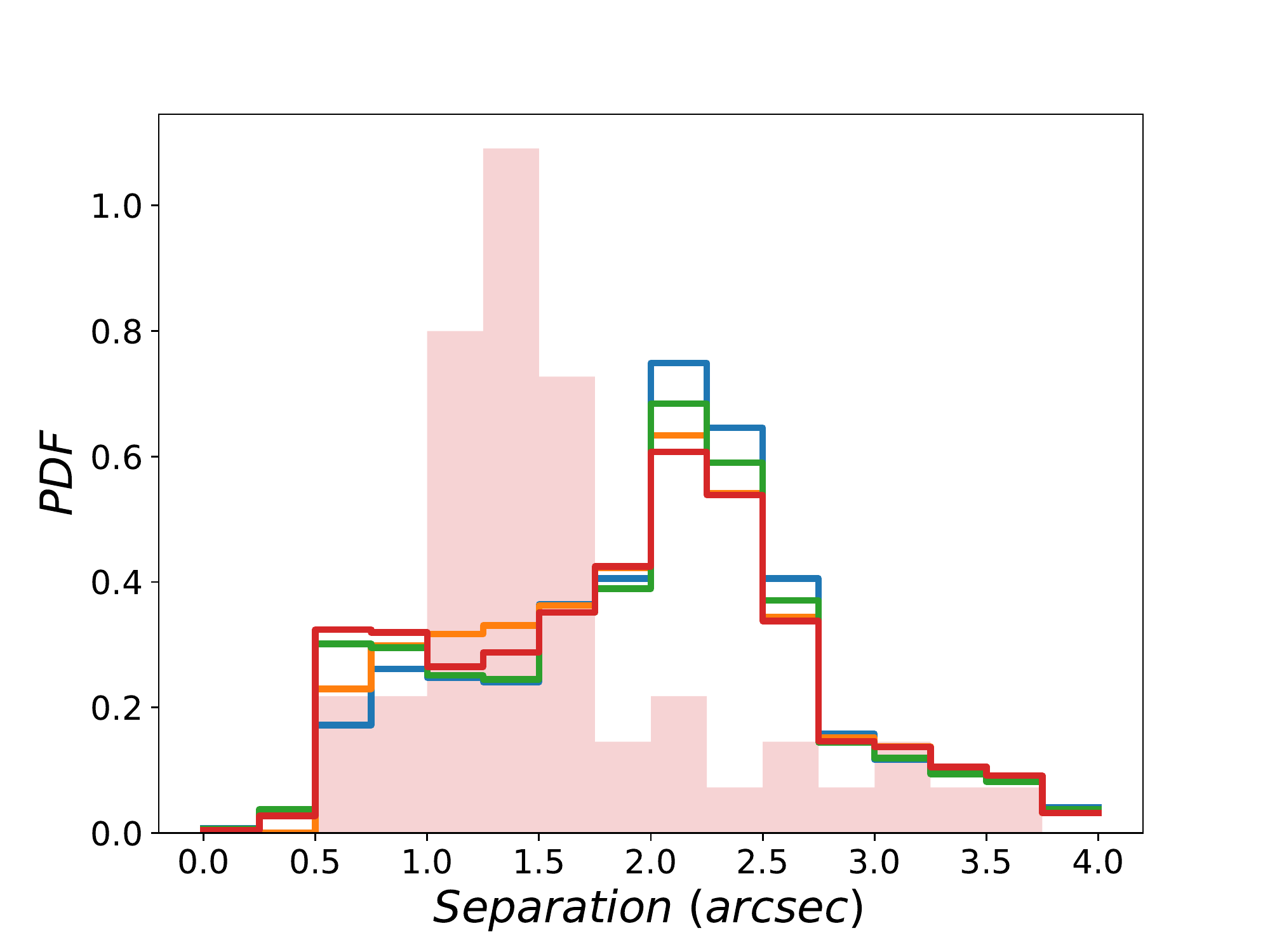}
        \caption{Comparisons between H22-new-RLQ, H22-new-D22, H22-new-L22, and H22-new of distributions of $S_{colour}$ (upper left), $S_{RF}$ (upper right), $g$-band magnitude (lower left), and image separation (lower right).}
        \label{fig:H22new}
    \end{figure*}

We have identified 620 new candidates not found in any of RLQ, D22, or L22. We name the catalogue of these new candidates H22-new, which can be achieved by requiring `in\_RLQ'=False, `in\_D22'=False, and `in\_L22'=False simultaneously in the online catalogue. In Fig.\,\ref{fig:H22new}, we compare the distributions of $S_{colour}$, $S_{RF}$, $g$-band magnitude, and image separation between H22-new and the new systems found in H22 but not RLQ (H22-new-RLQ), not D22 (H22-new-D22) and not L22 (H22-new-L22).

% These are labelled H22-new in the online catalogue.

Considering the distribution of $S_{colour}$, H22-new candidates have lower scores than the other samples, because the high-score systems are more likely to be rediscovered in other datasets, i.e., RLQ, D22, and L22. Similar trends are also visible in the distribution of $S_{RF}$. In terms of the $g$-band magnitude, H22-new candidates occupy a fainter region than the other samples; there are more matches with the other datasets at brighter fluxes due to higher signal-to-noise ratios. Regarding image separations, H22-new candidates have  larger separations on average compared to the other samples, meaning that the systems with small image separations are more likely to be rediscovered in the other datasets. 

In summary, H22-new is a sample of multiply imaged QSO candidates not found in any of RLQ, D22 or L22. Although some candidates with high confidence (included in D22 and L22) are excluded in H22-new, the 101 Grade-As in H22-new are still valuable candidates worth spectroscopic follow-up. H22-new also includes 214 Grade-Bs and 305 Grade-Cs. These samples contain more false positives than in Grade-As, but mining multiply imaged QSOs systems from them can still enhance the sample size of multiply imaged QSOs. Hence, spectroscopic confirmations are also worth pursuing for Grade-Bs and Cs, albeit at a lower priority.

    \subsection{Basic lens modelling of selected candidates}
    \label{sec:lens_modelling}
        
        Lens modelling provides a useful tool for enhancing the grading of candidate lensed QSO systems. A statistical goodness-of-fit of a lens model can be used to improve the confidence of a candidate being a confirmed lensed system. Thus we perform basic lens modelling of the candidates with visible lens light in H22-new-RLQ to validate our identification algorithm. As such, we select 52 candidates for modelling. To simplify the modelling, we choose the candidates with visible light centres and assume the mass centres align with the light centres. This reduces the number of model parameters. For example, doubly imaged QSOs have six independent data points, including the positions and fluxes of the two images, however, if we do not fix the centre of mass, there are eight free model parameters for the Singular Isothermal Ellipsoidal \citep[SIE, ][]{Kormann1994} model, including five lens parameters and three source parameters. 

        The mass model adopts the SIE profile, which has been widely applied to describe the mass distribution of galaxies \cite[e.g.,][]{Bolton12,Sonnenfeld13}. The SIE profile requires the five parameters ($x_l$, $y_l$, $q$, $phi$, $rein$), which are the position coordinates, the axis ratio, the position angle, and the Einstein radius, respectively. We fix the mass centre $x_l$, $y_l$ of each lens according to the peak of its light distribution as mentioned above. Three parameters are used to model the quasar source: two for the unlensed position ($\vec{r}_s$) and one for the flux ($f_{\rm s}$) on the source plane. We denote the observed position and flux of the $i-$th image of a lensed QSO by $\vec{r}_i$ and $f_i$, respectively. In our modelling, we fit to both the image fluxes and the image positions by varying the lens and source parameters. Quantitatively, we vary the model positions ($\vec{r}_i^{P}$) and fluxes ($M_i^{P} f_{\rm s}$) of the $i-$th image until the best fit with the corresponding observed parameters is found, i.e., when $\vec{r}_i^{P} \approx \vec{r}_i$ and $M_i^{P} f_{\rm s} \approx f_i$. Here $M_i^{P}$ is the model magnification factor at the $i-$th lensed image. In practice, the above objective is achieved by using a simulated annealing algorithm that minimises the following penalty function,

    \begin{equation}
    \begin{aligned}
    \chi^2 &= \chi_{position}^{2} + \chi_{flux}^{2} \\
    &= \sum_{i} \frac{\left|\vec{r}_{i}-\vec{r}_i^{P}\right|^{2}}{\sigma_i^2} + 
    \sum_{i} \frac{\left(f_{i}-M_{i}^{P} f_s\right)^{2}}{\sigma_{f, i}^{2}}
    \end{aligned}
    \label{eq:h0_td_5}
    \end{equation}

    where $\sigma_i$ and $\sigma_{f, i}$ are the position and flux measurement uncertainties at the $i-$th image, respectively. The calculations of $\vec{r}_i^P$ and $M_i^P$ are performed with the open-source software {\tt lenstronomy} \citep{Birrer2015, Birrer2018, Birrer2021}, and the python code used for the lensing analysis in this work is publicly available\footnote{\url{https://github.com/caoxiaoyue/model_lensed_quasar}}. To evaluate the goodness of fit, we adopt the Bayesian information criterion \citep[BIC, ][]{Liddle2007}, given by

    \begin{equation}
        BIC=\ln(n)k-2\ln(\hat{L})
    \end{equation}

    where $\hat{L}=\exp(-\chi_{minimised}^2)$; $n$ is the number of input data points and $k$ is the number of free parameters of the model. For pairs, $n=k=6$, while for the two quads (with internal IDs of 110720 and 2484493), $n=9,k=6$, since our approach missed one of the quadruply lensed images as shown in the first two panels of Fig.\ref{fig:cands}. A lower BIC indicates a better agreement between the model and the observations.
    
\begin{figure*}
        \centering
        \includegraphics[scale=0.2]{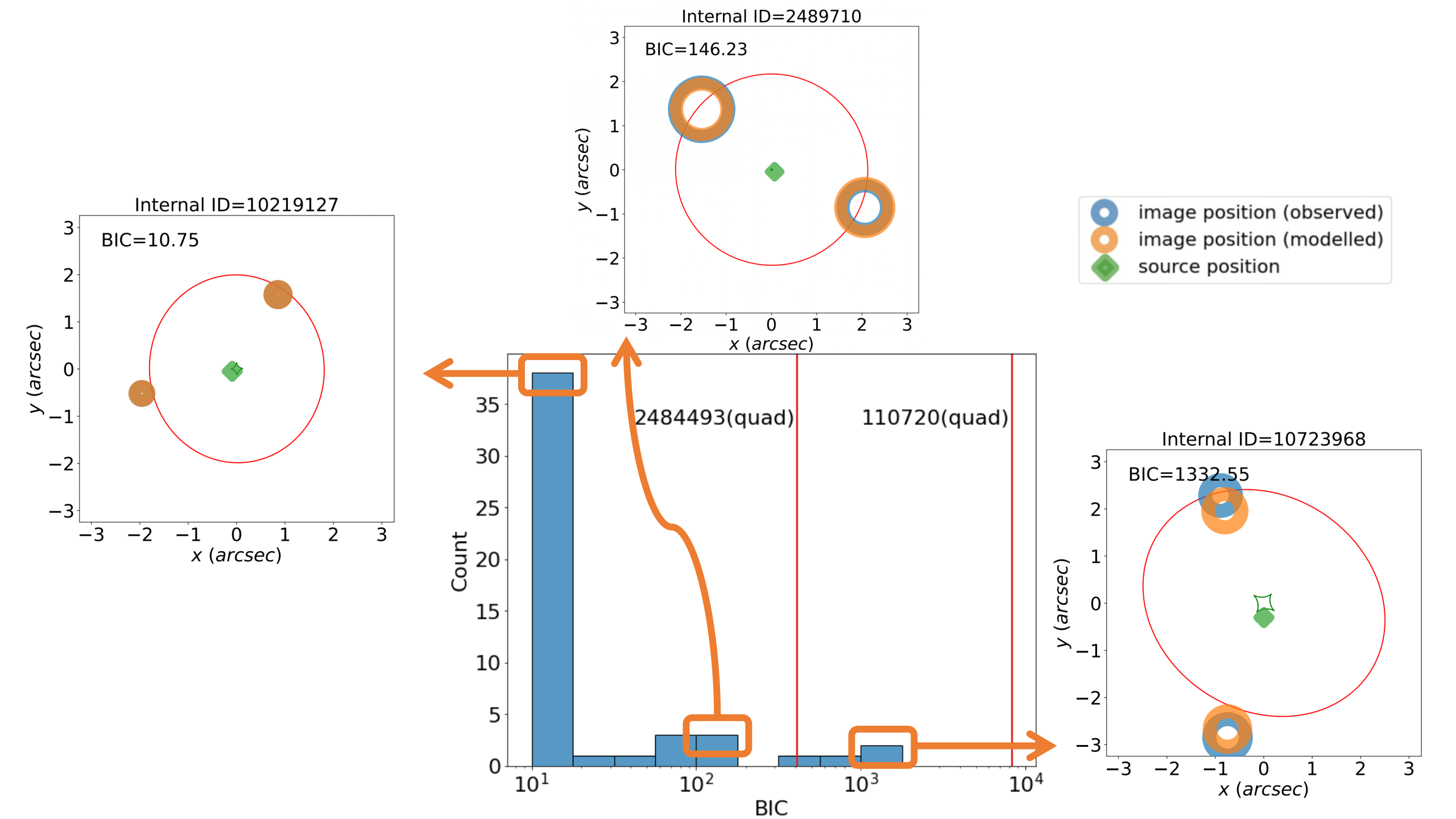}
        \caption{The distribution of BICs of the 52 lensing models. BICs of two-image systems are represented by histograms, while vertical orange lines indicate the BICs of the two quad systems. In the square panels, blue circles show the observed image positions, orange circles show the modelled image positions, and green diamonds show the source positions given by the lens models. The green and red curves indicate the caustic and critical curves of the corresponding lensing models.}
        \label{fig:BIC}
    \end{figure*}

    Consequently, we determine lens parameters ($q$,$phi$,$rein$) of 52 systems, including 50 pairs and 2 quads. The distribution of BICs of the above models is shown in Fig.\ref{fig:BIC}. Examples of three different systems with different BICs are also plotted in Fig.\ref{fig:BIC} to demonstrate the agreement between the best-fit models and observations. There is no significant disagreement between models and observations, even for the two quad systems with a larger BIC. The mean BIC of Grade-A, B, and C two-image samples are 14.94, 18.47, and 135.12, respectively, which suggests that the outcomes of our human grading procedure are in good agreement with the modelling.
    
\begin{figure*}
     \centering
        \includegraphics[scale=0.29]{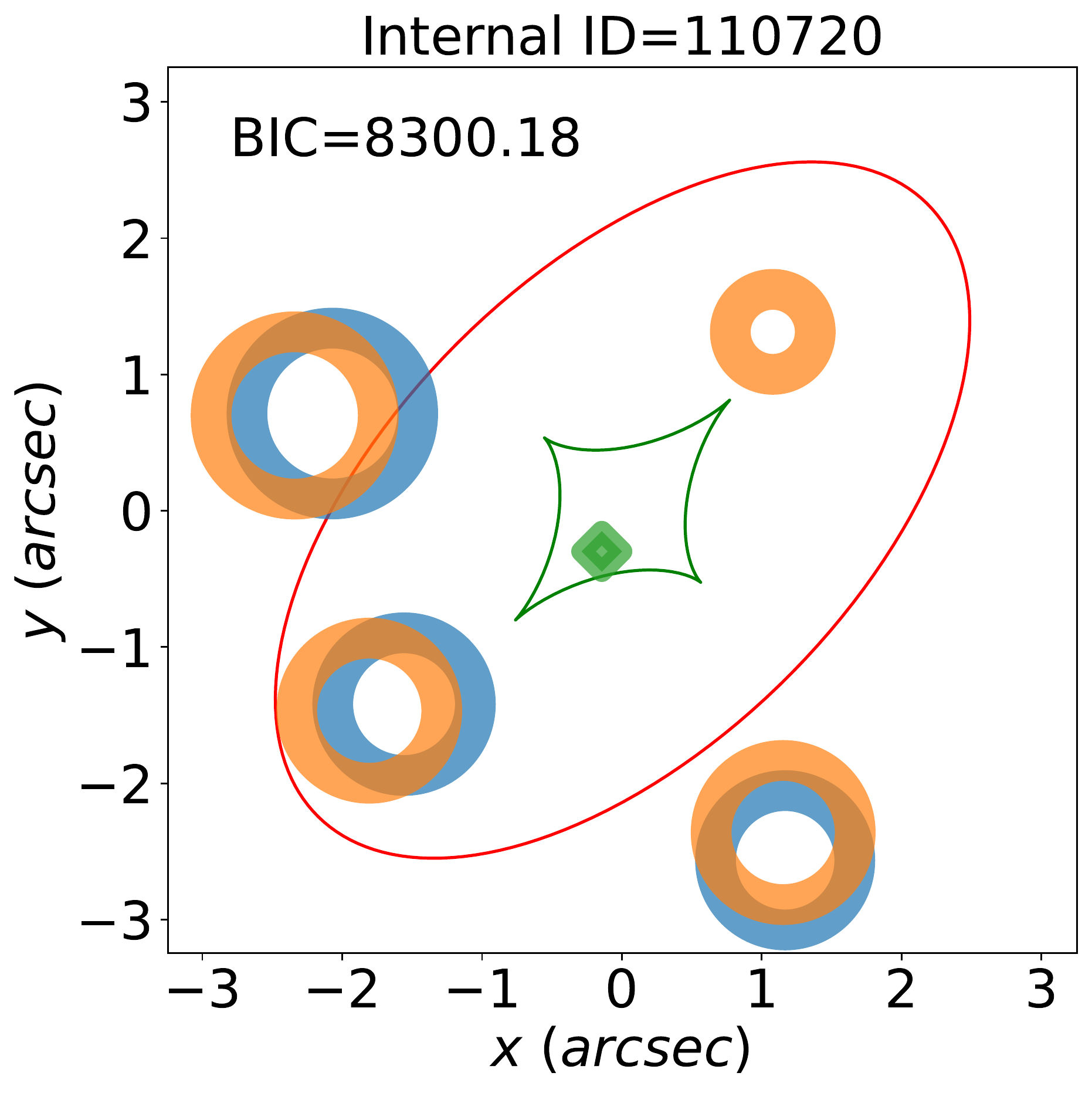}
        \includegraphics[scale=0.29]{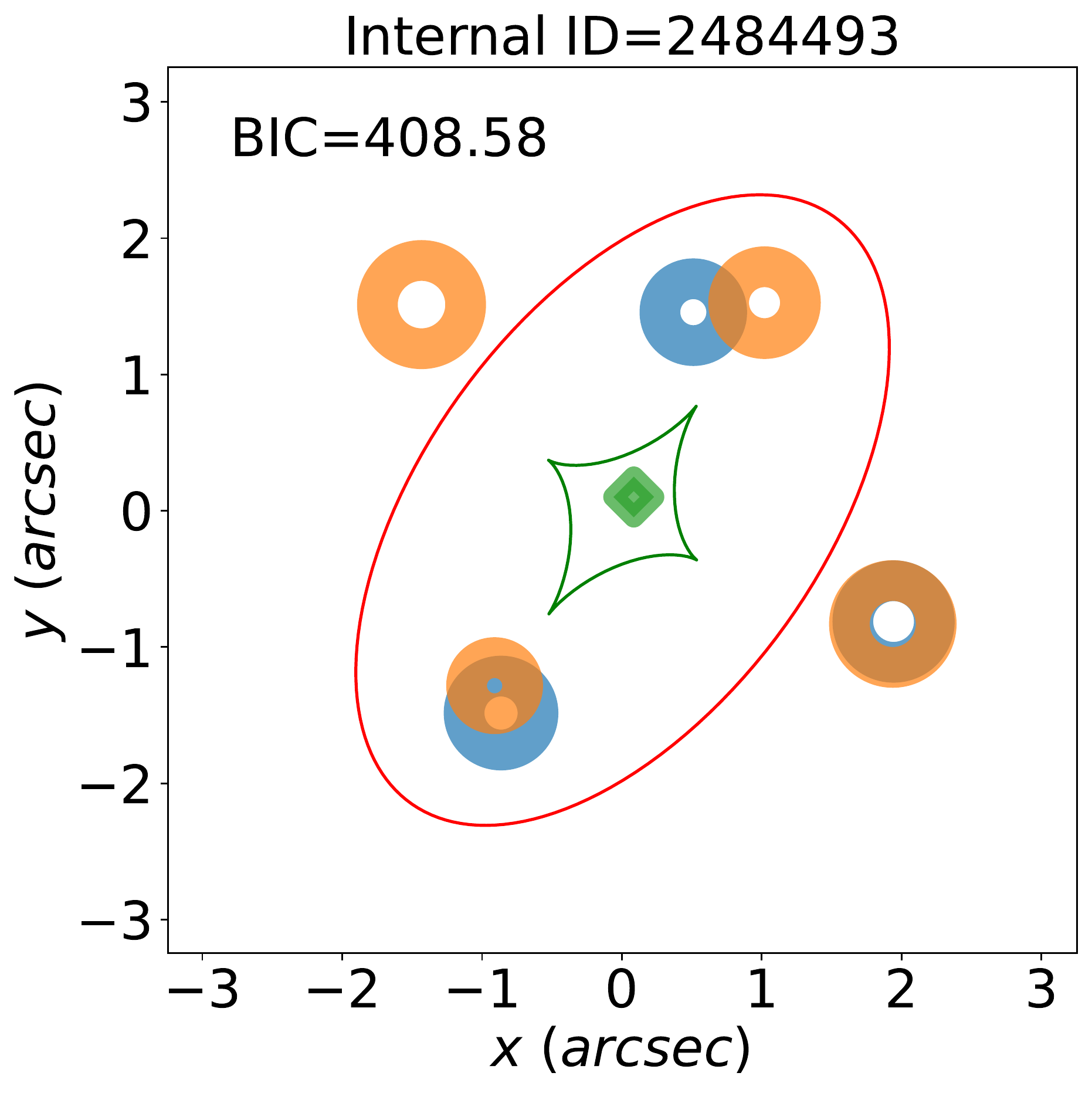}
        \includegraphics[scale=0.29]{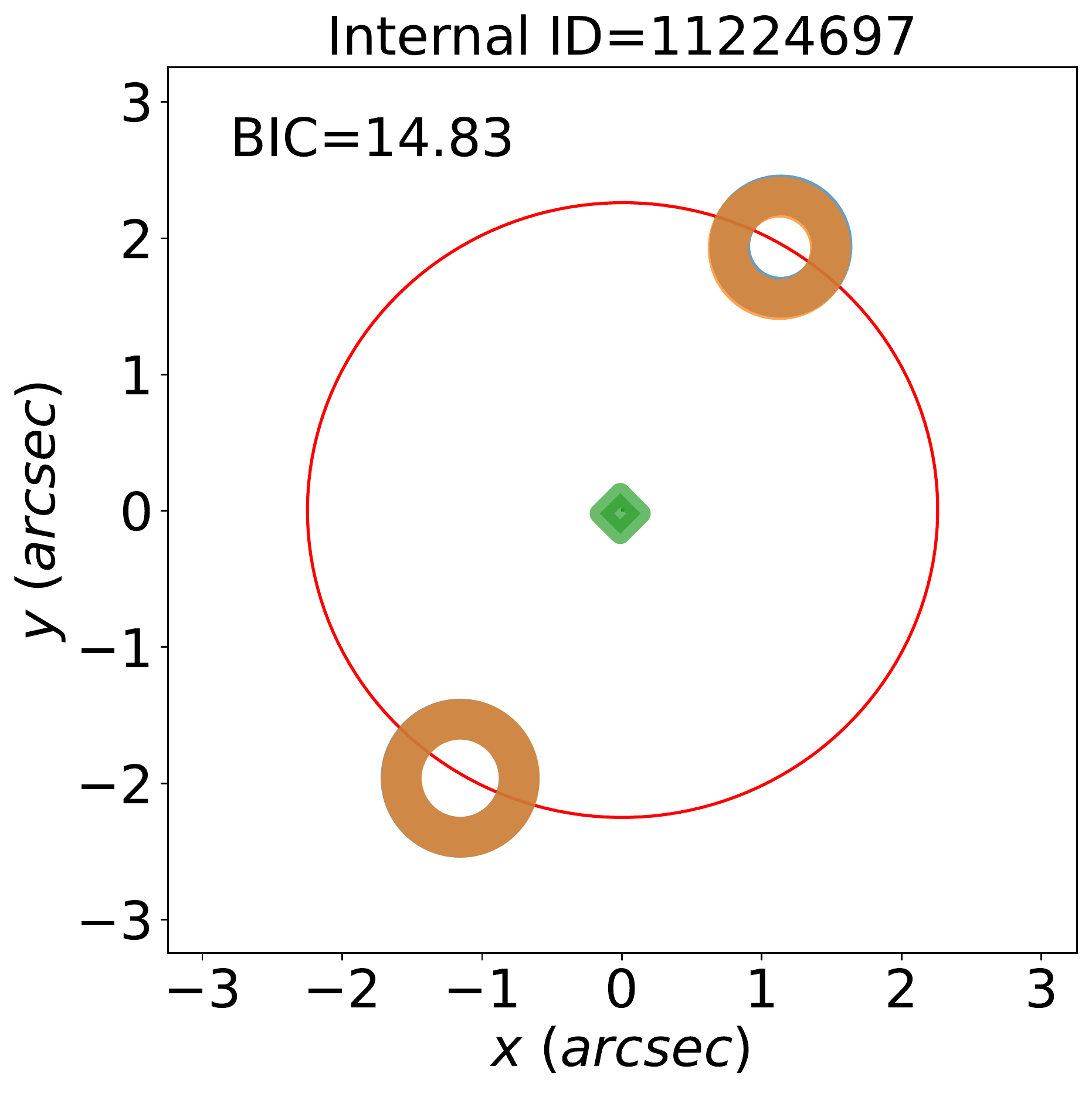}
        \caption{The modelled caustic curves for the three candidates in the third line of Fig.\,\ref{fig:cands}. The plots are centred at the lens light centre, which is determined by peak of the lens light. Symbols are the same as in Fig.\,\ref{fig:BIC}.}
    \label{fig:caustic}
    \end{figure*}

    We also present three sample systems in the bottom panels in Fig.\,\ref{fig:caustic}, including one double-image system and two quadruple-image systems. Notably, the system with internal ID 110720 \citep[also reported by][as DESI-055.7976-28.4777]{Huang2021}, which has the largest BIC, seems to be a triple-image system, but the best-fit model presents a quad system that leads to the "worst" fitting case in this work. The primary reason is that the parent catalogue probably misses the fourth image due to the brightness limit or morphology classification. Hence, the quick lens modelling process can decrease the candidate sample's false positive rates and improve the integrity of individual systems.

\section{Discussion and Conclusions}
\label{sec:conclusion}
    In this work, we have designed a catalogue-based approach for mining multiply imaged lensed QSO candidates from large catalogue datasets. The approach adopts three stages: 

    \begin{enumerate}
        \item identifying groups of candidate QSOs; 
        \item refining the groups using a statistical measure of the similarity of group member colours and the likelihood of members being QSOs according to \cite{He2022}. This includes both removal of entire groups and removal of individual members for groups with more than two members.
%        \item refining the groups, separately dealing with two different situations: 
%        \begin{enumerate}
%        \item for the groups that contain two members, selecting the groups successively according to the colour similarity of the group members $S_{colour}$ and the likelihood of group members being QSOs $S_{RF}$ given by \cite{He2022};
%        \item for the groups with more than two members, cleaning the outlier members successively according to the $S_{RF}$ and $S_{colour}$;
%        \end{enumerate}
        \item employing human inspection to grade candidate systems and further reject unlikely systems.
    \end{enumerate}
    
We have applied the approach to the catalogue of \cite{He2022} which comprises 24,440,816 QSO candidates created from the DESI-LS DR9 source catalogue. Initially, 562,206 QSO candidate groups were identified, further refined by colour similarity and QSO likelihood into a set of 102,468 candidate groups containing two members and a set of 45,905 candidate groups containing more than two members. Subsequent human inspection resulted in a final total of 971 candidate multiply imaged QSOs. Of these, 620 candidates are new in the sense that they have not been identified in existing studies. Relevant images and catalogues of our candidate systems are available online\footnote{\url{https://github.com/EigenHermit/lensed\_qso\_cand\_catalogue\_He-22}}.

The redshift range of the 971 candidate lensed QSOs is $(0, 3.5]$, peaking at $\sim 2$. Their $g$-band apparent magnitudes  span the range $17 \leq g \leq 24$, peaking at $g \sim 20.5$. We have estimated the recovery rate of our mining from the 57 known lensed QSOs contained in the  catalogue of \cite{He2022}; our catalogue of 971 contains 53 of these, indicating a recovery rate of approximately $93\%.$ We note that the use of these 57 known systems in optimising our mining strategy likely results in selection bias towards discovering similar systems. Therefore, this recovery rate only approximately represents the true recovery rate which will be lower in practice.

Our candidate catalogue includes grades awarded by two independent human inspectors, according to how likely the candidates are actually lensed QSO systems. This grading is based on the visual inspection of images, relying mainly on image configurations but also turning to colour similarity and QSO likelihood in unclear cases. The grading process is subjective, and so the inclusion of both awarded grades in our catalogue gives an indication of subjectivity. There are 284 Grade-As, 283 Grade-Bs, and 404 Grade-Cs. Grade-As show high similarity to the distribution of separations of the multiply imaged QSOs systems in OM10, while Grade-Bs and Grade-Cs generally have larger separations. We also implement lens modelling on selected candidates to validate the candidates, and the results show reasonable outcomes, especially for Grade-A systems. Hence, we consider the Grade-A systems as prioritised targets for follow-up spectroscopic campaigns. 
        
Compared to the studies of D22 and L22, ours adopts a catalogue-based approach and starts with a larger parent sample. In particular, our parent sample is deeper than those of D22 and L22. D22 utilises a relatively shallow parent sample ($r<22.7$), and about $2/3$ of D22 candidates have been identified in our catalogue of 971 candidates. The remaining $\sim 1/3$ is caused by different parent samples (contributing about $68\%$ difference) and selection methodologies (contributing about $32\%$ difference). Our method also applies colour similarity thresholding, which improves purity but sacrifices completeness; for example, two out of the 57 known systems are rejected after our selection based on colour similarity. There are 44 systems in common between H22 and L22. Among them, 39 candidates are confirmed lenses or promising candidates, and 5 are false positives. 

Since our method applies directly to catalogue data, it is heavily dependent on the source extraction algorithm used which may not be optimised for finding strongly lensed QSO systems. Hence, we are generally biased against finding small image separation systems and quadruply imaged systems. This is true of the DESI-LS catalogue mined in this work. The PSF of DESI-LS results in about $64.0\%$ of quads (or $41.5\%$ of pairs) being missed when creating the initial list of QSO candidates because the quads are more likely to be smeared to one point source. Although we ensured a high recovery rate of known lenses, the selection by colour and QSO likelihood results in some lensed QSO being rejected, as is shown in comparison with D22. Due to this process, 26 Grade-A candidates (according to D22's grading) were missed. As such, in future work, we plan to develop a methodology that combines catalogue-based and image-based approaches to take advantage of both whilst avoiding their disadvantages.

To summarise, our work provides the largest catalogue of multiply imaged lensed QSO candidates to date, comprising 620 new lensed QSO candidates of which over 100 are high-grade. With future large-scale spectroscopic follow-up of these from, for example, DESI and the 4-metre Multi-Object Spectrograph Telescope \citep{deJong2019}, existing samples of lensed QSOs used for cosmological and astrophysical studies could therefore be greatly increased in size for much-improved statistical power.

\begin{acknowledgements}

We thank China-VO for providing DESI-LS data. We thank Huanyuan Shan, Xinzhong Er, Dongxu Zhang, Yunfei Chen for the extensive discussions. We thank \textbf{\textit{astropy}}, \textbf{\textit{healpix}}, \textbf{\textit{pandas}}, \textbf{\textit{seaborn}}, \textbf{\textit{lenstronomy}} for providing convenient and reliable python packages. We acknowledge the support from 1. the Ministry of Science and Technology of China (Nos. 2020SKA0110100); 2. the science research grants from the China Manned Space Project (No. CMS-CSST-2021-A01); 3. CAS Project for Young Scientists in Basic Research (No. YSBR-062). HZ acknowledge the supports by the National Natural Science Foundation of China (NSFC, Grant No. 12120101003),  Beijing Municipal Natural Science Foundation under grant 1222028 and the science research grants from the China Manned Space Project with No. CMS-CSST-2021-A02.

\\

This project used data obtained with the Dark Energy Camera (DECam), which was constructed by the DES collaboration. Funding for the DES Projects has been provided by the U.S. Department of Energy, the U.S. National Science Foundation, the Ministry of Science and Education of Spain, the Science and Technology Facilities Council of the United Kingdom, the Higher Education Funding Council for England, the National Center for Supercomputing Applications at the University of Illinois at Urbana-Champaign, the Kavli Institute of Cosmological Physics at the University of Chicago, Center for Cosmology and Astro-Particle Physics at the Ohio State University, the Mitchell Institute for Fundamental Physics and Astronomy at Texas A\&M University, Financiadora de Estudos e Projetos, Fundacao Carlos Chagas Filho de Amparo, Financiadora de Estudos e Projetos, Fundacao Carlos Chagas Filho de Amparo a Pesquisa do Estado do Rio de Janeiro, Conselho Nacional de Desenvolvimento Cientifico e Tecnologico and the Ministerio da Ciencia, Tecnologia e Inovacao, the Deutsche Forschungsgemeinschaft and the Collaborating Institutions in the Dark Energy Survey. The Collaborating Institutions are Argonne National Laboratory, the University of California at Santa Cruz, the University of Cambridge, Centro de Investigaciones Energeticas, Medioambientales y Tecnologicas-Madrid, the University of Chicago, University College London, the DES-Brazil Consortium, the University of Edinburgh, the Eidgenossische Technische Hochschule (ETH) Zurich, Fermi National Accelerator Laboratory, the University of Illinois at Urbana-Champaign, the Institut de Ciencies de l’Espai (IEEC/CSIC), the Institut de Fisica d’Altes Energies, Lawrence Berkeley National Laboratory, the Ludwig Maximilians Universitat Munchen and the associated Excellence Cluster Universe, the University of Michigan, NSF’s NOIRLab, the University of Nottingham, the Ohio State University, the University of Pennsylvania, the University of Portsmouth, SLAC National Accelerator Laboratory, Stanford University, the University of Sussex, and Texas A\&M University.
\end{acknowledgements}

\bibliographystyle{aa}
\bibliography{cite}

\end{document}